\documentclass[aps,prl,reprint,superscriptaddress,amsmath,amssymb,floatfix,nobibnotes,nofootinbib]{revtex4-2}
\usepackage{graphicx}
\usepackage{xcolor}
\usepackage{hyperref}
\hypersetup{colorlinks=true,citecolor=blue,linkcolor=blue,urlcolor=blue}
\newcommand{\dd}{\mathrm{d}}
\newcommand{\smfigwidth}{3.4in}
\newcommand{\smmultifigwidth}{5.0in}
\newtheorem{proposition}{Proposition}
\makeatletter
\renewcommand{\@begintheorem}[2]{\trivlist\item[\hskip\labelsep{\bfseries #1\ #2.}]\itshape}
\makeatother

\begin{document}
\title{Aubry--Andr\'e transition in photonic time quasicrystals}
\author{Shuhang Chen}
\email{chenshuhang@mail.ustc.edu.cn}
\affiliation{Department of Electronic Engineering and Information Science, University of Science and Technology of China, Hefei, Anhui 230027, China}
\author{Qi Zhu}
\email{zhuqi@ustc.edu.cn}
\affiliation{Department of Electronic Engineering and Information Science, University of Science and Technology of China, Hefei, Anhui 230027, China}
\begin{abstract}
In one dimension, Anderson localization by a disordered potential sets in at any strength, whereas localization by a quasiperiodic potential in the Aubry--Andr\'e model occurs through a transition at a critical strength. Disordered photonic time crystals exhibit a temporal counterpart of this Anderson localization, in which light, instead of decaying in space, stops propagating and grows in time. Whether a counterpart of the Aubry--Andr\'e transition exists in photonic time quasicrystals remains an open question. Here we show that it does, and that the transition occurs at mobility edges in momentum rather than in energy. These edges arise even under impulsive modulation, where the field obeys the self-dual almost-Mathieu equation exactly, and for general analytic modulations Avila's global theory reveals the condition under which they form. The Aubry--Andr\'e transition thus extends from spatial lattices to the time domain, where it sorts light by wave number into components that propagate and components that stop and grow.
\end{abstract}
\maketitle
A photonic time crystal is a spatially uniform medium whose permittivity is modulated periodically in time~\cite{Lustig2018,Galiffi2022}. Uniformity conserves momentum while the modulation exchanges energy with the wave, so band gaps open in momentum rather than in frequency~\cite{Zurita-Sanchez2009,Reyes-Ayona2015}. Inside such a gap the wave is amplified parametrically, the mechanism behind amplified emission, lasing, and free-electron radiation in time-varying media~\cite{Lyubarov2022,Dikopoltsev2022,Lee2026PRL}, at a frequency pinned by the period of the modulation.

\begin{figure}[!t]
\centering
\includegraphics[width=\columnwidth]{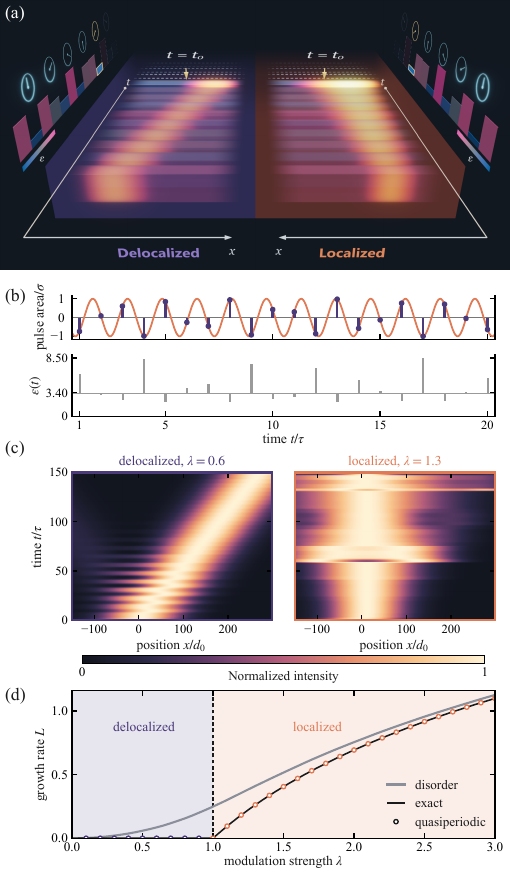}
\caption{Photonic time quasicrystal and its localization transition. (a) Spatially uniform medium at successive times. (b) Pulse areas, proportional to $\cos X_m$ (top), and the resulting permittivity (bottom); the pulse width is schematic. (c) Wave-packet intensity at $kd_0=\pi/2$ in the delocalized ($\lambda=0.6$, $L=0$) and localized ($\lambda=1.3$, $L=\ln1.3$) phases. Each time row is normalized, because the localized intensity grows by more than 30 orders of magnitude. (d) Growth rate $L$ versus $\lambda$ at $E=0$: exact (black), quasiperiodic (circles), and disordered (gray) pulses.}\label{fig1}
\end{figure}

Without periodicity, nothing pins the frequency, and amplification need not occur inside a gap. Disorder is the extreme case: when the permittivity fluctuates randomly from one interval to the next, the energy of a wave grows exponentially, as derived for random temporal modulation~\cite{Carminati2021} and observed with water waves~\cite{Apffel2022}, while the group velocity of a pulse decays exponentially to zero~\cite{Sharabi2021}. Temporal disorder thus localizes waves in the time domain~\cite{Sharabi2021,Eswaran2025,Sacha2016,Delande2017}, with ensemble statistics that, as shown in Ref.~\cite{Sharabi2021}, are those of Anderson localization~\cite{Anderson1958,Lagendijk2009}. Just as for its spatial counterpart in one dimension~\cite{Billy2008,Lahini2008}, this localization sets in at any strength of the disorder and at every wave number~\cite{Furstenberg1963,TessieriIzrailev1999,MallickMarcq2002}.

Quasiperiodic lattices change this. In the Aubry--Andr\'e model, waves localize only when the amplitude of the potential exceeds a critical value~\cite{AubryAndre1980}. Self-duality fixes this value exactly~\cite{AubryAndre1980,Jitomirskaya1999}, and the transition has been observed with ultracold atoms~\cite{Roati2008} and in photonic lattices~\cite{Lahini2009}. For potentials without self-duality, localization generally sets in at mobility edges in energy~\cite{BiddleDasSarma2010,Ganeshan2015,Luschen2018,An2021}. Avila's global theory then replaces self-duality with a quantized slope of the Lyapunov exponent in the complexified phase~\cite{Avila2015}, from which exact mobility edges have been derived for several families~\cite{Wang2020mosaic,WangXiaYouZhengZhou2023}. Quasiperiodicity can also be imposed in time. Kicking atomic matter waves in this way realizes a three-dimensional Anderson transition, through incommensurate frequencies that act as additional dimensions~\cite{Casati1989,Chabe2008}. For light in a photonic time quasicrystal, the momentum gaps form fractal sets~\cite{Ruchiev2026,Liu2025OL,Coppolaro2025,Araujo2025} or a Hofstadter butterfly~\cite{Ni2025}, but whether temporal localization occurs through a counterpart of the Aubry--Andr\'e transition remains an open question~\cite{Sharabi2021,NemirovskyLevy2023}.

Here we show that photonic time quasicrystals host this transition. Below a critical modulation strength, light propagates or grows at locked frequencies in gaps; above it, light stops and grows at unlocked frequencies in a localized phase. For impulsive modulation, we find that each wave number $k$ obeys the almost-Mathieu equation with a pulse strength that grows with $k$, so its self-dual point becomes a mobility edge in momentum, and the position of this edge can be controlled by shaping the pulses. For general analytic modulation, which lacks self-duality, we identify the transition by Avila's acceleration~\cite{Avila2015}, a topological integer. The integer is obtained by shifting the modulation's phase variable into the complex plane, a mathematical construction that corresponds physically to gain and loss and has served to classify both localization and amplification in non-Hermitian quasicrystals~\cite{Longhi2019,LiuZhouChen2021,Chen2026NHPTQC}. Here it identifies the localized phase by a jump from zero to a positive value. The triad that organizes wave transport in space, from Bloch bands through Anderson localization to the Aubry--Andr\'e transition, thereby closes in the time domain~\cite{Bloch1929,Anderson1958,AubryAndre1980}.

\begin{figure*}[!t]
\includegraphics[width=0.75\textwidth]{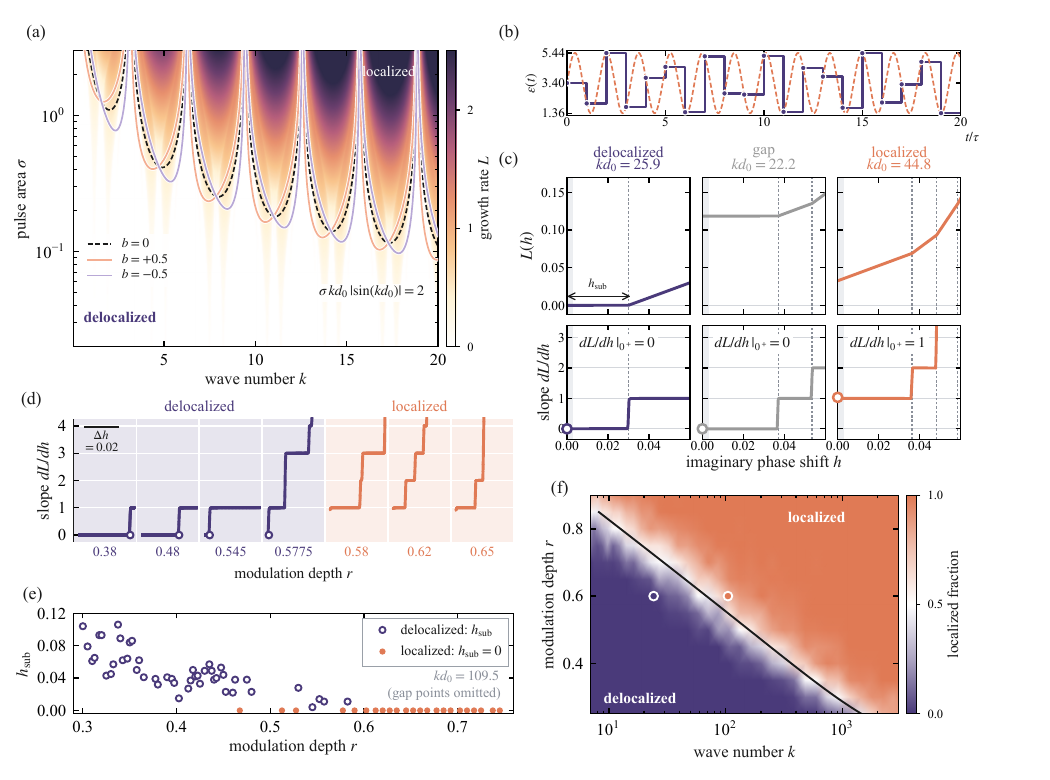}
\caption{Mobility edge in momentum and acceleration criterion. (a) Growth rate $L$ of the impulsive model in the plane of wave number $k$ (in units of $1/d_0$) and pulse area $\sigma$; exact mobility edges for $b=0$ (dashed) and $b=\pm0.5$ (colored); the narrow tongues below them are momentum gaps. (b) Permittivity $\epsilon(t)$ of the layered crystal at $r=0.6$ for $f(X)=\sin X$. (c) $L(h)$ and $dL/dh$ at $r=0.7$ for delocalized ($kd_0=25.9$), gap ($22.2$), and localized ($44.8$) wave numbers, where $h$ is the imaginary phase shift and $h_{\rm sub}$ is the half-width of the zero-exponent strip; the slopes at $h\to0^+$ are $0,0,1$, with $N=50000$. (d) At $kd_0=109.5$, $dL/dh$ versus $h$ for increasing $r$; the slope at $h\to0^+$ jumps from 0 to 1 between $r=0.5775$ and $0.58$. (e) $h_{\rm sub}$ versus $r$ at $kd_0=109.5$; it decreases with $r$ and vanishes where localized points appear, near $r=0.47$, and no delocalized point remains above $r=0.5825$. Each $r$ samples a different spectral point at fixed $kd_0$, producing scatter; gap points are omitted. (f) Share of localized states among the delocalized and localized ones, gaps excluded; the black line is the mobility edge, a smooth fit through logistic midpoints, and the circles select the two wave-number bands in Fig.~\ref{fig3}. Purple, gray, and orange denote delocalized, gap, and localized throughout.}\label{fig2}
\end{figure*}

A layered photonic time quasicrystal~\cite{Chen2026NHPTQC} is a spatially uniform medium whose permittivity is constant within each layer in time and follows a quasiperiodic sequence from layer to layer [Fig.~\ref{fig1}(a)],
\begin{equation}
 \epsilon_m=\epsilon_R[1+r f(X_m)],\qquad X_m=2\pi(\alpha m+\phi),
 \label{eq:model}
\end{equation}
where $\epsilon_m$ is the permittivity of layer $m$, $\epsilon_R$ is the reference permittivity, $r$ is the modulation depth, $f$ is a real, $2\pi$-periodic analytic function, $\alpha$ is an irrational number (the golden mean $(\sqrt5-1)/2$ throughout), and $\phi\in[0,1)$ is an initial phase. All layers have the same duration $\tau$, and the permittivity changes abruptly at layer boundaries, as in the layered descriptions of periodic and disordered time crystals~\cite{Lustig2018,Sharabi2021}.

Because the medium is uniform in space, the wave number $k$ is conserved, and each $k$ is governed by its own one-dimensional problem in time, with no coupling between different $k$. We describe the field by the state vector $\mathbf v=(D,\dot D/\omega_0)^T$, where $D$ is the amplitude of the electric displacement $\mathbf D$, $\dot D/\omega_0$ is proportional to the amplitude of the magnetic induction $\mathbf B$, and $\omega_0=ck/\sqrt{\epsilon_R}$ is the reference frequency. Both components are continuous across each boundary, so the state vector evolves layer by layer as
\begin{equation}
 \begin{gathered}
 \mathbf v_{m+1}=A(X_m)\mathbf v_m,\\
 A(X_m)=\begin{pmatrix}\cos kd_m&(d_0/d_m)\sin kd_m\\-(d_m/d_0)\sin kd_m&\cos kd_m\end{pmatrix},
 \end{gathered}
 \label{eq:matrix}
\end{equation}
where $d_m=c\tau/\sqrt{\epsilon_m}$ is the distance that light travels in layer $m$ and $d_0=c\tau/\sqrt{\epsilon_R}$ is its value in the reference medium, so that $kd_m$ is the phase accumulated in layer $m$. Because $d_m/d_0$ does not depend on $k$, the wave number enters only through $kd_0$, and we quote wave numbers in units of $1/d_0$.

When the modulation is concentrated into short pulses, the dynamics at each wave number reduces exactly to the Aubry--Andr\'e model. In this limit the permittivity equals $\epsilon_R$ except during pulses separated by $\tau$, which we treat as instantaneous [Fig.~\ref{fig1}(b)]. Integrating the wave equation across pulse $m$ leaves $D$ continuous and changes $\dot D/\omega_0$ by $-kd_0\,\sigma\cos X_m\,D$, where $\sigma\cos X_m$ is the pulse area, the time integral of the change in inverse permittivity in units of $\tau/\epsilon_R$ (SM Sec.~S2), and $\sigma\geq0$; a positive area lowers the permittivity and a negative one raises it. One period then maps the state vector as
\begin{equation}
 \mathbf v_{m+1}=\begin{pmatrix}\cos kd_0&\sin kd_0\\-\sin kd_0&\cos kd_0\end{pmatrix}\begin{pmatrix}1&0\\-kd_0\,\sigma\cos X_m&1\end{pmatrix}\mathbf v_m.
 \label{eq:pulse}
\end{equation}

\begin{figure}[!t]
\includegraphics[width=\columnwidth]{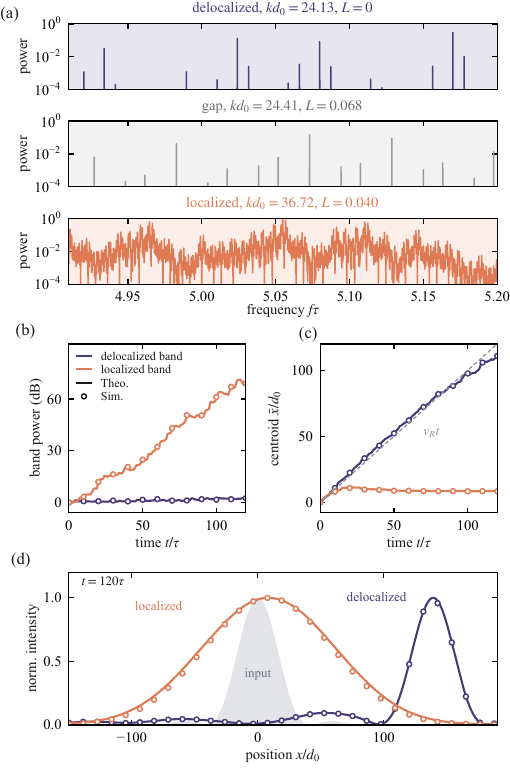}
\caption{Frequency spectra and wave-number filtering. (a) Power spectra of the field at three fixed wave numbers for $r=0.6$, computed after factoring out the overall exponential growth and normalized to their maxima: delocalized ($kd_0=24.13$), gap ($24.41$), and localized ($36.72$). (b) Band power relative to input; (c) band centroids $\bar x$ relative to the launch center, including reflected tails; (d) final intensity profiles, separately normalized, with the input shaded. Position $x$ and time $t$ are scaled by $d_0$ and $\tau$, respectively. The input bands, at $r=0.6$, are $kd_0\in[24.05,24.21]$ and $[104.33,104.49]$. Theo.\ (lines): transfer matrices; Sim.\ (circles): staggered-grid Yee-FDTD calculation.}\label{fig3}
\end{figure}

Eliminating $\dot D/\omega_0$ from Eq.~(\ref{eq:pulse}), and shifting $X_m$ by $\pi$ when $\sin kd_0<0$, gives the almost-Mathieu equation
\begin{equation}
 D_{m+1}+D_{m-1}+2\lambda\cos X_m\,D_m=E\,D_m,
 \label{eq:amo}
\end{equation}
where the energy $E=2\cos kd_0$ and the modulation strength $\lambda=\sigma kd_0|\sin kd_0|/2$ are both set by the wave number. Its self-duality, which exchanges $(E,\lambda)$ with $(E/\lambda,1/\lambda)$, places the Aubry--Andr\'e transition at the critical modulation strength $\lambda=1$~\cite{AubryAndre1980}. The two phases are distinguished by the growth rate $L$, which is the Lyapunov exponent of the product of transfer matrices averaged over the initial phase $\phi$, so that after $m$ periods the field amplitude grows as $e^{Lm}$. On the spectrum, $L=\max(0,\ln\lambda)$~\cite{BourgainJitomirskaya2002,Avila2015} (SM Sec.~S2). $L=0$ marks the delocalized phase ($\lambda<1$), where light keeps propagating, and $L>0$ the localized phase ($\lambda>1$), where light stops and grows [Fig.~\ref{fig1}(c)].

The wave packets in Fig.~\ref{fig1}(c) show this contrast directly. At $\lambda=0.6$ the packet translates, whereas at $\lambda=1.3$ it stays at the launch region and broadens there. To follow $L$ across the transition without entering a gap, we fix $kd_0=\pi/2$, where $E=0$ lies in the spectrum for every $\lambda$, and vary the modulation strength (pulse area); the computed $L$ follows the exact result [Fig.~\ref{fig1}(d)]. With disordered kicks, obtained by drawing the phases $X_m$ independently with the same single-pulse distribution, the same recurrence instead describes a linear kicked oscillator, whose parametric instability is Anderson localization~\cite{TessieriIzrailev1999}, and $L$ is positive at every $\lambda$ (gray). Disorder therefore localizes at any modulation strength, whereas quasiperiodic order localizes only beyond the transition at $\lambda=1$.

Both $E$ and $\lambda$ depend on the wave number, and $\lambda$ carries the factor $kd_0$, because at fixed pulse area the jump in $\dot D/\omega_0$ is proportional to $kd_0$. The critical point $\lambda=1$ therefore becomes the curve
\begin{equation}
 \sigma=\frac{2}{kd_0|\sin kd_0|}
 \label{eq:pulseboundary}
\end{equation}
in the plane of wave number and pulse area [Fig.~\ref{fig2}(a)]. In the spatial almost-Mathieu operator (the Aubry--Andr\'e--Harper model), all energies share one modulation strength, whereas here each wave number carries its own, so the transition occurs at a mobility edge in momentum. For any $\sigma>0$, wave numbers near $kd_0=(n+\tfrac12)\pi$ are localized once $kd_0>2/\sigma$, while those near $kd_0=n\pi$, where $\lambda\to0$, always remain delocalized; the narrow tongues of positive growth rate below the curve are momentum gaps.

Shaping the pulses moves this edge. For example, pulse areas $\sigma\cos X_m/(1-b\cos X_m)$ with $|b|<1$ turn Eq.~(\ref{eq:amo}) into the generalized Aubry--Andr\'e model~\cite{Ganeshan2015}, and the edge is then known exactly (SM Sec.~S2). A positive $b$ lowers the critical pulse area on the low-wave-number side of each lobe of Eq.~(\ref{eq:pulseboundary}) and raises it on the high side, a negative $b$ does the opposite, and all edges cross at $kd_0=(n+\tfrac12)\pi$ [Fig.~\ref{fig2}(a)].

For a general analytic $f(\cdot)$, the layered quasicrystal of Eq.~(\ref{eq:matrix}) has no such self-duality. This raises the question of whether the Aubry--Andr\'e transition survives without it, and how the transition can then be located. We take $f(X)=\sin X$ with $0<r<1$ as an example [Fig.~\ref{fig2}(b)] and continue $X$ to $X+ih$, with an imaginary shift $h$ inside the strip $|h|<\operatorname{arccosh}(1/r)$ where $\epsilon(X+ih)$ has no zero. Physically, this continuation adds gain and loss modulated in quadrature with the permittivity~\cite{Chen2026NHPTQC}; mathematically, it is the central step of Avila's global theory~\cite{Avila2015}, which reads the transition from the response of the growth rate to $h$.

Let $L(h)$ be this growth rate, computed with $X_m$ replaced by $X_m+ih$ at fixed $k$ and $r$, so that $L(0)=L$. The theory makes $L(h)$ even, convex, and piecewise affine with integer slopes, so that its slope at $h\to0^+$,
\begin{equation}
 \nu=\left.\partial_h^+L(h)\right|_{h=0},
 \label{eq:acceleration}
\end{equation}
called the acceleration~\cite{Avila2015}, is a nonnegative integer, tied to the winding number that the continued problem accumulates as the phason $\phi$ runs once around its circle~\cite{LiuZhouChen2021}.

The acceleration sorts wave numbers into three classes [Fig.~\ref{fig2}(c)]. In the delocalized phase, $L(h)$ vanishes on a strip $|h|<h_{\rm sub}$, so that $L(0)=\nu=0$. In a momentum gap, $L(h)$ stays flat at $L(0)>0$, so that $\nu=0$~\cite{Avila2015}, and light grows at a frequency that the rotation number locks to a gap label~\cite{AvilaJitomirskaya2010}. In the localized phase, $L(h)$ rises from $h\to0^+$, so that $L(0)>0$ and $\nu\geq1$, and light grows at unlocked frequencies. The gaps are the quasiperiodic counterparts of those in periodic photonic time crystals and appear as the resonance tongues of quasiperiodic Hill equations~\cite{BroerPuigSimo2003}.

Raising the modulation depth at a fixed wave number shows how localization sets in. At $kd_0=109.5$, the first step of $dL/dh$ moves toward $h=0$ as $r$ grows, and once it reaches $h=0$ the slope at $h\to0^+$ jumps from 0 to 1 [Fig.~\ref{fig2}(d)]. The half-width $h_{\rm sub}$ measures how much gain and loss a delocalized wave number tolerates without growing, and localization sets in where this tolerance closes. The statistics of $h_{\rm sub}$ over many values of $r$ show this closing [Fig.~\ref{fig2}(e)]. Because each $r$ places the fixed wave number at a different point of the spectrum, delocalized and localized points interleave between $r\approx0.47$ and $0.58$ before the localized phase takes over.

Mapping this classification over the whole plane of wave number and modulation depth turns the crossover into a mobility edge in momentum. Figure~\ref{fig2}(f) shows the share of localized wave numbers among the delocalized and localized ones, with gaps excluded. Delocalized wave numbers dominate at small $kd_0$ and localized ones at large $kd_0$, and the edge between them, taken where the two shares are equal (SM Sec.~S5), retreats to high wave numbers as the modulation weakens, from $kd_0\approx70$ at $r=0.6$ to a scaling $r^{-2.2}$ at weak modulation. Unlike in the spatial Aubry--Andr\'e model, a single sinusoidal harmonic thus suffices for a mobility edge, and the edge divides any pulse by wave number: components below it propagate, while those above it stop and grow.

The edge retreats because the two processes within a layer respond differently to the wave number. At each boundary, temporal reflection mixes forward and backward waves by an amount set by the permittivity ratio alone; within the layer, the field accumulates the phase $kd_m$, which varies from layer to layer by an amount of order $kd_0r$ for any $f(\cdot)$. Only the phase variation grows with $k$, and this intuition can be made exact. A conjugation separates the two processes and turns Eq.~(\ref{eq:matrix}) into
\begin{equation}
 \tilde A_m=\begin{pmatrix}e^{-\eta_m}&0\\0&e^{\eta_m}\end{pmatrix}\begin{pmatrix}\cos kd_m&\sin kd_m\\-\sin kd_m&\cos kd_m\end{pmatrix},
 \label{eq:normal}
\end{equation}
where the diagonal factor, with $\eta_m=\tfrac14\ln(\epsilon_{m+1}/\epsilon_m)$, describes the reflection. When $\alpha$ is Diophantine, as the golden mean is, the modulation never comes close to repeating, and a further conjugation absorbs the phase variation, leaving a constant rotation times a perturbation of size $r\,e^{O(kd_0r)}$ (SM Sec.~S6). A constant rotation has zero acceleration, and small analytic perturbations cannot raise it~\cite{Avila2015,JKS2009}, so localization requires
\begin{equation}
 kd_0\geq\frac{c_1}{r}\ln\frac{c_2}{r}\quad(r\to0),
 \label{eq:lower}
\end{equation}
with positive constants $c_1$ and $c_2$ that depend only on $\alpha$ and $f(\cdot)$ (SM Sec.~S6).

The numerically determined edge, which scales as $r^{-2.2}$, lies above this bound, as it must. At any fixed wave number, Eq.~(\ref{eq:lower}) sets a finite modulation depth below which localization is impossible, whereas random ordering of the same layers localizes every wave number at any depth~\cite{Furstenberg1963}. The contrast that separates Aubry--Andr\'e from Anderson localization therefore needs no self-duality and holds for every analytic modulation of a photonic time quasicrystal.

The three phases can be told apart in the frequency spectrum of the field at a single wave number [Fig.~\ref{fig3}(a)]. The permittivity switches once per layer and the modulation phase $X_m$ grows by $2\pi\alpha$ from one layer to the next, so the crystal shifts the frequency of light by integer multiples of $1/\tau$ and $\alpha/\tau$. In the delocalized phase, a change of basis turns the transfer matrices into a constant rotation by $2\pi\rho$, where $\rho$ is the rotation number, and the bounded field contains only the frequencies $f\tau=\pm\rho+n\alpha+j$ with integers $n$ and $j$ (SM Sec.~S3).
 Its spectrum is a discrete comb. In a gap the comb persists with $\rho$ locked to a gap label, and only the growth of the field sets it apart. In the localized phase the growing field spreads its power over a continuous band. As in the Aubry--Andr\'e duality between position and momentum~\cite{AubryAndre1980}, a field extended in time has a discrete spectrum, whereas a localized field has a continuous one.

The division of a pulse by the mobility edge can be followed directly in space. We launch a pulse containing two equal-power bands, centered at the delocalized wave number of Fig.~\ref{fig3}(a) and at the localized wave number $kd_0=104.41$, and propagate it with the transfer-matrix theory (Theo.) and with an independent full-wave Yee-FDTD calculation (Sim.), which agree throughout [Figs.~\ref{fig3}(b)--\ref{fig3}(d)]. After 120 layers the localized band has gained about $68$~dB, while the delocalized band has gained little [Fig.~\ref{fig3}(b)]. Over the same time, the delocalized band travels at the speed of light in the reference medium, whereas the localized band remains near the launch point [Fig.~\ref{fig3}(c)]. The final profiles show the delocalized band intact apart from a weak tail left behind by temporal reflection, and the localized band broadened symmetrically about the launch point [Fig.~\ref{fig3}(d)]. The band that grows is thus the band that stops.

Temporal reflection ties the arrest to the growth. Spatial uniformity conserves momentum, so the difference between the forward and backward power in a band stays at its input value $P(0)$, while the modulation pumps energy into their sum $P(t)$. For a narrow band, the group velocity, the speed of the intensity centroid once the modulation is switched off, therefore drops as $v_g=v_RP(0)/P(t)$, with $v_R=c/\sqrt{\epsilon_R}$ (SM Sec.~S7). The energy gained is thus shared almost equally by forward and backward waves. The $68$-dB gain of the localized band brings $v_g$ down to about $10^{-7}v_R$. Arrest accompanying growth has been observed in disordered photonic time crystals~\cite{Sharabi2021}; here it acts selectively in wave number.

Removing the phase variation tests this mechanism directly. Choosing layer durations $\tau_m=\tau\sqrt{\epsilon_m/\epsilon_R}$, as in the equal-phase construction of Ref.~\cite{Ni2025}, makes every layer accumulate the same phase $kd_0$, so the rotation in Eq.~(\ref{eq:normal}) no longer varies from layer to layer and the product stays within $O(r)$ of a constant rotation at every wave number. Equal-phase modulation then leaves every wave number delocalized or in a gap, rigorously at weak modulation, where the argument behind Eq.~(\ref{eq:lower}) now holds uniformly in $k$, and numerically up to a permittivity contrast of about $2\times10^{3}$ (SM Fig.~S8). The transition reported here thus rests on two ingredients, an analytic modulation, for which the acceleration is defined, and an accumulated phase that varies from layer to layer. Earlier photonic time quasicrystals differ in one of the two: two-valued substitution sequences~\cite{Ruchiev2026,Liu2025OL,Coppolaro2025,Araujo2025} are not samples of an analytic function on the circle, so the acceleration is undefined for them, and the equal-phase construction removes the phase variation.

Observing the transition requires modulation that is abrupt enough and strong enough. Each boundary reflects only if the permittivity switches fast compared with the wave period, so a finite switching time suppresses temporal reflection at high wave numbers and confines the localized phase below a cutoff wave number that decreases as the switching time grows (SM Sec.~S8). The edge must therefore lie below this cutoff. For $r\gtrsim0.6$, a permittivity contrast of about four, the edge lies at $kd_0\approx10$--$100$ [Fig.~\ref{fig2}(f)], and about a hundred layers already separate the two bands by tens of decibels [Fig.~\ref{fig3}(b)]. Switched transmission lines, in which temporal reflection and broadband frequency translation have been observed~\cite{Moussa2023}, switch faster than the wave period and offer a route to this regime. Their output spectrum would reveal the transition directly, as a discrete comb below the edge and a continuum above it [Fig.~\ref{fig3}(a)]. Because each wave number reduces to an oscillator with a piecewise-constant frequency, the same transition occurs in any wave medium switched quasiperiodically in time~\cite{Apffel2022,ElasticBeams2025,KimChongDaraio2024,HungGurarieChin2013}.

Quasiperiodic time modulation thus turns temporal localization into an Aubry--Andr\'e transition, which sets in only above a critical modulation strength and occurs at mobility edges in momentum. Under impulsive modulation, the self-dual point of the almost-Mathieu equation becomes this edge. Under analytic modulation, where self-duality is absent, Avila's acceleration locates the edge, and a rigorous bound pushes it to high wave numbers at weak modulation because the phase accumulated in each layer varies more strongly as the wave number grows. What separates Aubry--Andr\'e from Anderson localization is therefore not self-duality. The transition splits light by wave number into a part that propagates and a part that stops and grows, and at a single wave number it appears as a frequency comb on one side and a continuum on the other, which is Aubry--Andr\'e duality read in time and frequency. Because time modulation exchanges the roles of energy and momentum, the spectral theory of quasicrystals carries over to time-varying media with wave number as its parameter~\cite{Zurita-Sanchez2009,Lustig2018,Galiffi2022}. Each wave number also behaves as an oscillator with a time-dependent frequency, a setting in which time-varying media can create photon pairs from vacuum~\cite{Lyubarov2022,SustaetaOsuna2025}, so the transition may offer a way to shape quantum light in momentum. A similar approach could apply to other systems driven quasiperiodically in time, such as parametric oscillators and atomic condensates~\cite{HungGurarieChin2013}.

\input{ms.bbl}

\clearpage
\onecolumngrid
\setcounter{section}{0}
\setcounter{figure}{0}
\setcounter{table}{0}
\setcounter{equation}{0}
\renewcommand{\thesection}{S\arabic{section}}
\renewcommand{\thetable}{S\arabic{table}}
\renewcommand{\thefigure}{S\arabic{figure}}
\renewcommand{\theequation}{S\arabic{equation}}
\renewcommand{\theHsection}{S\arabic{section}}
\renewcommand{\theHfigure}{S\arabic{figure}}
\renewcommand{\theHtable}{S\arabic{table}}
\renewcommand{\theHequation}{S\arabic{equation}}
\begin{center}
{\large\bfseries Supplemental Material\par}
\vspace{8pt}
Shuhang Chen$^{1,*}$ and Qi Zhu$^{1,\dagger}$\par
\vspace{3pt}
{\small\itshape $^1$Department of Electronic Engineering and Information Science,\\
University of Science and Technology of China, Hefei, Anhui 230027, China\par}
\end{center}
\makeatletter
\onecolumn@grid@setup
\let\set@footnotewidth\set@footnotewidth@one
\let\compose@footnotes\compose@footnotes@one
\makeatother
\begingroup
\renewcommand{\thefootnote}{\fnsymbol{footnote}}
\footnotetext[1]{\href{mailto:chenshuhang@mail.ustc.edu.cn}{chenshuhang@mail.ustc.edu.cn}}
\footnotetext[2]{\href{mailto:zhuqi@ustc.edu.cn}{zhuqi@ustc.edu.cn}}
\endgroup
This Supplemental Material follows the order of the main text. Wave numbers are quoted in units of $1/d_0$, where $d_0=c\tau/\sqrt{\epsilon_R}$ is the distance light travels in one layer of the reference medium, so the dimensionless wave number is $kd_0$. We write $\omega_0=ck/\sqrt{\epsilon_R}$, $v_R=c/\sqrt{\epsilon_R}$, and $X_m=2\pi(\alpha m+\phi)$, with $\alpha=(\sqrt5-1)/2$ unless stated otherwise. The growth rate $L$ is the Lyapunov exponent per layer averaged over the initial phase $\phi$. $L(h)$ is its value with $X$ replaced by $X+ih$, $\nu=\partial_h^+L(0)$ is the acceleration, and $h_{\rm sub}$ is the half-width of the strip on which $L(h)=0$.

\section{Layer transfer matrix}
\label{sec:sm_matrix}

Consider a spatially uniform, nondispersive medium with constant permeability
$\mu$ and permittivity $\epsilon_0\epsilon(t)$, where $t$ denotes time. For a transverse plane wave propagating along the spatial coordinate $x$,
$D=D_k(t)e^{ikx}\hat y$ and $B=B_k(t)e^{ikx}\hat z$, where $D_k$ and $B_k$ are the electric-displacement and magnetic-induction amplitudes and $\hat y,\hat z$ are transverse unit vectors. A dot denotes a time derivative. Maxwell's equations give
\begin{equation}
 \dot D_k=-\frac{ik}{\mu}B_k,\qquad
 \dot B_k=-\frac{ik}{\epsilon_0\epsilon(t)}D_k .
 \label{eq:sm_maxwell_first}
\end{equation}
Eliminating $B_k$ gives
\begin{equation}
 \ddot D_k+\frac{c^2k^2}{\epsilon(t)}D_k=0,\qquad
 c^2=(\epsilon_0\mu)^{-1}.
 \label{eq:sm_wave}
\end{equation}
The wave number is conserved because the medium is uniform. At a temporal
interface, $D$ and $B$ remain continuous in the absence of surface sources;
Eq.~(\ref{eq:sm_maxwell_first}) then makes $\dot D_k$ continuous as well.

Inside layer $m$, $\epsilon=\epsilon_m$, with
$d_m=c\tau/\sqrt{\epsilon_m}$ and $d_0=c\tau/\sqrt{\epsilon_R}$. Define $z(X)=d_m/d_0=\sqrt{\epsilon_R/\epsilon(X)}$; the phase
accumulated in one layer is $kd_m$. Starting at the layer boundary $t_m=m\tau$, with elapsed time $t'$, the solution is
\begin{equation}
 D(t_m+t')=D(t_m)\cos[(ck/\sqrt{\epsilon_m})t']
 +\frac{\dot D(t_m)}{ck/\sqrt{\epsilon_m}}\sin[(ck/\sqrt{\epsilon_m})t'].
 \label{eq:sm_layer_solution}
\end{equation}
For equal-duration layers we use $\epsilon(X)=\epsilon_R[1+r\sin X]$ and
$X_m=2\pi(\alpha m+\phi)$, with $\alpha=(\sqrt5-1)/2$. For
$\mathbf v=(D,\dot D/\omega_0)^{\mathsf T}$,
\begin{equation}
 \begin{gathered}
 A(X)=\begin{pmatrix}\cos kd_m&(d_0/d_m)\sin kd_m\\-(d_m/d_0)\sin kd_m&\cos kd_m\end{pmatrix}
      =S(X)^{-1}R(kd_m)S(X),\\
 S(X)=\begin{pmatrix}\sqrt{d_m/d_0}&0\\0&\sqrt{d_0/d_m}\end{pmatrix},\qquad
 R(\theta)=\begin{pmatrix}\cos\theta&\sin\theta\\-\sin\theta&\cos\theta\end{pmatrix}.
 \end{gathered}
 \label{eq:sm_layer}
\end{equation}
$\det A=1$.
Since $\dot D/\omega_0=-ikB/(\mu\omega_0)$, the two state components are
proportional to the continuous Maxwell fields $D$ and $B$.
Section~\ref{sec:sm_normal} uses this factorization to separate interface mixing from the accumulated phase. Each layer preserves
$(d_m/d_0)D^2+(d_0/d_m)(\dot D/\omega_0)^2$, proportional to the field energy by a factor that is constant within that
layer. The entries of $A$ depend on $X$ through
$(d_m/d_0)^2=[1+r\sin X]^{-1}$ and are analytic for
$|\operatorname{Im}X|<h_{\rm sing}=\operatorname{arccosh}(1/r)$.

\subsection{Forward and backward waves}

In layer $m$ define
$b_\pm=\tfrac12(D\pm i\dot D/(ck/\sqrt{\epsilon_m}))$, so
$D=b_+e^{-ick/\sqrt{\epsilon_m}t}+b_-e^{ick/\sqrt{\epsilon_m}t}$. Continuity at an interface from
$\epsilon$ to $\epsilon'$, with frequencies $\omega=ck/\sqrt{\epsilon}$ and $\omega'=ck/\sqrt{\epsilon'}$ before and after the interface, gives
\begin{equation}
 \begin{pmatrix}b_+'\\ b_-'\end{pmatrix}
 =\frac12\begin{pmatrix}1+\omega/\omega'&1-\omega/\omega'\\
 1-\omega/\omega'&1+\omega/\omega'\end{pmatrix}
 \begin{pmatrix}b_+\\ b_-\end{pmatrix}.
 \label{eq:sm_fresnel}
\end{equation}
With $\omega/\omega'=\sqrt{\epsilon'/\epsilon}=e^{2\eta}$ and
$\eta=\tfrac14\ln(\epsilon'/\epsilon)$, the temporal transmission and reflection amplitude coefficients are
\begin{equation}
 T=e^\eta\cosh\eta,\qquad R=-e^\eta\sinh\eta,\qquad R/T=-\tanh\eta .
 \label{eq:sm_fresnel_coeff}
\end{equation}
In the reference normalization $a_\pm=\tfrac12(D\pm i\dot D/\omega_0)$,
$T_c=\tfrac12\begin{pmatrix}1&i\\1&-i\end{pmatrix}$ changes to the circular basis. With the boost $H(\eta)=\operatorname{diag}(e^{-\eta},e^{\eta})$, it gives
\begin{equation}
 T_cR(\theta)T_c^{-1}=\operatorname{diag}(e^{-i\theta},e^{i\theta}),\qquad
 T_cH(\eta)T_c^{-1}=
 \begin{pmatrix}\cosh\eta&-\sinh\eta\\-\sinh\eta&\cosh\eta\end{pmatrix}.
 \label{eq:sm_circular_map}
\end{equation}
The interface is a boost of rapidity $\eta$ between forward and backward waves;
the determinant-one boost and rotation factors lie in $SU(1,1)$.
In a fixed reference basis the real determinant-one layer maps preserve
$|a_+|^2-|a_-|^2$.

\section{Impulsive modulation}
\label{sec:sm_impulsive}

Write $\epsilon(t)^{-1}=\epsilon_R^{-1}+\delta(\epsilon^{-1})(t)$.
For the $m$th pulse define the dimensionless signed area by
\begin{equation}
 \frac{\epsilon_R}{\tau}\int_{\text{pulse }m}
 \delta(\epsilon^{-1})(t)\,\dd t=\sigma\cos X_m,
 \qquad \sigma\geq0.
 \label{eq:sm_area}
\end{equation}
Writing $P=\dot D/\omega_0$, integrating Eq.~(\ref{eq:sm_wave}) gives
\begin{equation}
 \Delta\dot D=-\omega_0^2\int D(t)\,\dd t
 -c^2k^2\int\delta(\epsilon^{-1})(t)D(t)\,\dd t.
 \label{eq:sm_pulse_integral}
\end{equation}
In the limit of fixed area and vanishing duration, $D(t)$ can be replaced by
$D(t_m)$ in the second integral. The first integral vanishes and
$\Delta P=-kd_0\sigma\cos X_mD(t_m)$, while $\Delta D$ vanishes.
For a pulse of duration $\delta t$, this approximation requires
$\delta t/\tau\ll1$, $\omega_0\delta t\ll1$, and
$\omega_0\delta t\,|kd_0\sigma\cos X_m|\ll1$, with a bounded pulse
profile at fixed area. In the normalized state, the omitted matrix terms are
$O[\omega_0\delta t(1+|kd_0\sigma\cos X_m|)^2]$ at fixed strength.
Positive area increases the inverse permittivity and lowers the permittivity;
negative area has the opposite sign. Negative areas require the permittivity to rise during the pulse, so a fixed negative area is reached only as a limit of finite pulses, which we propagate with Eq.~(\ref{eq:sm_finite_pulse}).

For pulse area $\sigma\cos X_m$, one period is
\begin{equation}
 \mathbf v_{m+1}=R(kd_0)J_m\mathbf v_m,\qquad
 J_m=\begin{pmatrix}1&0\\-kd_0\,\sigma\cos X_m&1\end{pmatrix}.
 \label{eq:sm_pulse}
\end{equation}
Writing $\widetilde P_m=P_m-kd_0\,\sigma\cos X_mD_m$, the forward and inverse free
steps give $D_{m+1}=D_m\cos(kd_0)+\widetilde P_m\sin(kd_0)$ and
$D_{m-1}=D_m\cos(kd_0)-P_m\sin(kd_0)$. Therefore
\begin{equation}
 D_{m+1}+D_{m-1}
 =\left[2\cos(kd_0)-\sigma kd_0\sin(kd_0)\cos X_m\right]D_m .
 \label{eq:sm_amo}
\end{equation}
After shifting $X$ by $\pi$ when $\sin(kd_0)<0$, this is the almost--Mathieu
equation with $E=2\cos(kd_0)$ and
$\lambda=\sigma kd_0|\sin(kd_0)|/2$. For $\sin(kd_0)\ne0$,
$(D_m,P_m)\mapsto(D_m,D_{m-1})$ is a constant invertible change of basis, so the
growth exponents agree. Explicitly, with $p=kd_0$,
\begin{equation}
 \begin{pmatrix}D_m\\D_{m-1}\end{pmatrix}
 =C_p\mathbf v_m,\qquad
 C_p=\begin{pmatrix}1&0\\\cos p&-\sin p\end{pmatrix},
 \qquad \det C_p=-\sin p.
 \label{eq:sm_amo_basis}
\end{equation}
The transformed step is $C_pR(p)J_mC_p^{-1}
=\bigl(\begin{smallmatrix}E-2\lambda\cos X_m&-1\\1&0\end{smallmatrix}\bigr)$,
after the stated phase shift. Products differ only by the fixed factors
$C_p,C_p^{-1}$, whose logarithmic norm contributions vanish after division by
$N$. This is the standard impulsively modulated oscillator/almost--Mathieu
correspondence~\cite{AubryAndre1980,TessieriIzrailev1999}.
At $\sin(kd_0)=0$, only the scalar recurrence loses the pulse term.
The full map is $(-1)^nJ_m$ and retains the shear; its growth need not be
bounded, although its exponential rate is zero for the bounded pulse sequence.

For irrational $\alpha$ and $E$ in the almost--Mathieu spectrum,
Corollary~2 of Ref.~\cite{BourgainJitomirskaya2002} gives $L(0)=\max(0,\ln\lambda)$.
Their potential is $\lambda_{\rm BJ}\cos X$, so
$\lambda_{\rm BJ}=2\lambda$ in our convention. Avila~\cite{Avila2015},
Appendix A gives the corresponding complexified formula
\begin{equation}
 L(h)=\max\{\max(0,\ln\lambda),\ \ln\lambda+|h|\}.
 \label{eq:sm_amo_lyap}
\end{equation}
Here Avila's imaginary shift is $\varepsilon=h/(2\pi)$.
The large-$h$ expansion
$2\lambda\cos(X+ih)=\lambda e^{h-iX}+O(e^{-h})$ gives
$L(h)=\ln\lambda+h+o(1)$. 

Zero energy lies in the spectrum for every $\lambda$, which justifies the choice $kd_0=\pi/2$ in Fig.~1(d). Let $H_{\phi,\lambda}$ denote the
almost--Mathieu operator. For a sequence $u=(u_m)$ on the integer lattice, the staggering map $(Uu)_m=(-1)^mu_m$ gives
$UH_{\phi,\lambda}U^{-1}=-H_{\phi+1/2,\lambda}$.
The spectrum and density-of-states measure are phase independent for
irrational $\alpha$, so both are symmetric under $E\mapsto-E$.
If zero lay in a gap, symmetry would give integrated density of states
$\mathcal N(0)=1/2$, where $\mathcal N(E)$ counts states below $E$ per lattice site. The gap-labeling theorem permits only
$\mathcal N(E)\in\alpha\mathbb Z+\mathbb Z$ in a gap
[Ref.~\cite{AvilaJitomirskaya2010}, Theorem 2.2]. For irrational $\alpha$, $1/2$ is not in this group.
This contradiction proves that zero belongs to the spectrum for every
$\lambda$; at $\lambda=0$ it also follows directly from the free band.

At fixed pulse area, $\lambda<1$ is met near $kd_0=n\pi$.
Writing $kd_0=n\pi+\delta$, where $n$ is an integer and $\delta$ is the detuning from $n\pi$, gives the leading envelope condition
$|\delta|<2/(\sigma kd_0)$ for $|\delta|\ll1$, and hence envelope width
$4/(\sigma kd_0)$ per period. This is a subcritical envelope, within which
individual momentum gaps can still occur.

The finite-width pulse, with relative permittivity $\epsilon_p$ and $d_p=c\tau/\sqrt{\epsilon_p}$, has its own exact matrix:
\begin{equation}
 A_p=\begin{pmatrix}\cos\theta_p&(d_0/d_p)\sin\theta_p\\
 -(d_p/d_0)\sin\theta_p&\cos\theta_p\end{pmatrix},\qquad
 \theta_p=(ck/\sqrt{\epsilon_p})\delta t .
 \label{eq:sm_finite_pulse}
\end{equation}
If $\delta t\to0$ with $kd_0\sigma=(ck/\sqrt{\epsilon_R})(d_p/d_0)^2\delta t$ fixed,
then
\begin{equation}
 A_p=\begin{pmatrix}1&0\\-kd_0\,\sigma&1\end{pmatrix}
 +O(\omega_0\delta t)+O[(1+|kd_0\sigma|)\theta_p^2].
 \label{eq:sm_impulse_limit}
\end{equation}

For one-sign pulse areas $\sigma_m=\sigma_0+\sigma\cos X_m$, where $\sigma_0$ is a constant offset, with
$|\sigma_0|\geq|\sigma|$, the constant part moves the auxiliary energy to
$E_{\rm eff}=2\cos(kd_0)-\sigma_0kd_0\sin(kd_0)$ and leaves $\lambda$ unchanged;
values of $E_{\rm eff}$ outside the spectrum are gaps.

For $\sigma kd_0\sin(kd_0)\ne0$, independent random phases satisfy
the Furstenberg conditions verified in Sec.~\ref{sec:sm_protocol}, so
$L>0$~\cite{Furstenberg1963}. The fibered rotation angle is
\begin{equation}
 \theta(kd_0)=\lim_{N\to\infty}\frac1N\arg^\#(\mathbf v_N),
 \label{eq:sm_rotation}
\end{equation}
where $\arg^\#$ is the continuous lift of the clockwise polar angle of
$\mathbf v_m=(D_m,\dot D_m/\omega_0)$. Free propagation gives $\theta=kd_0$, and a narrow packet moves at the group velocity $v_g=v_R\,\dd\theta/\dd(kd_0)$. For the delocalized packet of Fig.~1(c), $\dd\theta/\dd(kd_0)\approx1.8$, so its peak moves at about $1.8v_R$. Each pulse acts on the whole medium at the same instant and reshapes the packet everywhere at once, which allows the peak to move faster than $v_R$, while the energy flux in the reference medium never exceeds $v_R$ times the energy density.

\subsection{Shaped pulses}
\label{sec:sm_gaa}

The impulsive map also permits a controlled change of pulse shape.  Replace
$\cos X_m$ in Eq.~(\ref{eq:sm_amo}) by a profile with shape parameter $b$,
\begin{equation}
 g(X)=\frac{\cos X}{1-b\cos X},\qquad |b|<1 .
 \label{eq:sm_gaa_shape}
\end{equation}
With $X_m=2\pi(\alpha m+\phi)$ and propagation phase
$kd_0$, the recurrence is the generalized
Aubry--Andr\'e (GAA) equation
\begin{equation}
 D_{m+1}+D_{m-1}+\frac{2\lambda\cos X_m}{1-b\cos X_m}D_m=ED_m,
 \qquad E=2\cos(kd_0),\qquad
 \lambda=\frac{\sigma kd_0\sin(kd_0)}{2} .
 \label{eq:sm_gaa_map}
\end{equation}
The coupling is signed: for positive pulse area,
$\operatorname{sgn}\lambda=\operatorname{sgn}(\sin(kd_0))$.  Retaining this
sign is essential when the propagation phase crosses a half-period.

For unit hopping, the GAA self-duality relation is
\begin{equation}
 bE=2\operatorname{sgn}(\lambda)(1-|\lambda|),
 \label{eq:sm_gaa_duality}
\end{equation}
or, on the physical signed branch,
$bE+2\lambda=2\operatorname{sgn}(\lambda)$.  Hence the momentum-space
mobility edge is
\begin{equation}
 b\cos(kd_0)+\frac{\sigma kd_0\sin(kd_0)}{2}
 =\operatorname{sgn}(\sin(kd_0)),
 \label{eq:sm_gaa_edge}
\end{equation}
and for a positive pulse area
\begin{equation}
 \sigma_c(kd_0,b)=
 \frac{2\left[1-b\cos(kd_0)\,\operatorname{sgn}(\sin(kd_0))\right]}
 {kd_0|\sin(kd_0)|} .
 \label{eq:sm_gaa_ac}
\end{equation}
For $b=0$ this reduces to the cosine-pulse edge
$\sigma kd_0|\sin(kd_0)|=2$ used in the main text.

For energies in the GAA spectrum, the closed Lyapunov form is written with
$\chi=bE+2\lambda$:
\begin{equation}
 L(E)=\max\left\{0,\ln\!\left[
 \frac{H(\chi,b)}{1+\sqrt{1-b^2}}\right]\right\},
 \qquad
 H(\chi,b)=
 \begin{cases}
 \bigl(|\chi|+\sqrt{\chi^2-4b^2}\bigr)/2,& |\chi|\ge 2|b|,\\
 |b|,& |\chi|<2|b| .
 \end{cases}
 \label{eq:sm_gaa_lyap}
\end{equation}
To obtain this expression, let $S_E(X)$ be the determinant-one Schr\"odinger
cocycle and clear its pole with $D_E(X)=2[1-b\cos X]S_E(X)$.  For
$h_0=\operatorname{arcosh}(1/|b|)$, the distance to the nearest pole, Jensen's formula gives (with $\langle\cdot\rangle_X$ denoting the uniform average over $X\in[0,2\pi)$)
\begin{equation}
 \left\langle\ln\!\left|2[1-b\cos(X+ih)]\right|\right\rangle_X
 =\ln\!\left[1+\sqrt{1-b^2}\right],\qquad |h|<h_0 .
 \label{eq:sm_gaa_jensen}
\end{equation}
Thus the pole-cleared exponent differs from that of $S_E$ by this constant in
the analytic strip.  At large positive $h$,
\begin{equation}
 D_E(X+ih)=e^{h-iX}\left[
 \begin{pmatrix}-\chi&b\\-b&0\end{pmatrix}+o(1)\right],
 \qquad
 \rho_\pm=\frac{-\chi\pm\sqrt{\chi^2-4b^2}}{2} .
 \label{eq:sm_gaa_asymptote}
\end{equation}
The eigenvalues of the asymptotic constant matrix are $\rho_\pm$, and their leading modulus is $H(\chi,b)$; convexity of the complexified exponent and
the unit acceleration on the supercritical spectral branch then give
Eq.~(\ref{eq:sm_gaa_lyap}) after subtracting the Jensen term.  The spectral
condition excludes a positive, zero-acceleration branch, which would be
uniformly hyperbolic.  The
transition occurs at $H=1+\sqrt{1-b^2}$, or $|\chi|=2$, which gives
Eq.~(\ref{eq:sm_gaa_edge}) on the signed spectral branch.  The $b\to0$
limit is $L(E)=\max\{0,\ln|\lambda|\}$.  This is the
standard Avila global-theory construction~\cite{Avila2015}; the GAA spectral
formula and exact mobility edge are given in
Refs.~\cite{WangXiaYouZhengZhou2023,Ganeshan2015}, with related
complexification for a quasiperiodic mosaic model in Ref.~\cite{Wang2020mosaic}.
Direct transfer-matrix products agree with Eq.~(\ref{eq:sm_gaa_lyap}).

\begin{figure}[t]
\centering\includegraphics[width=\smmultifigwidth]{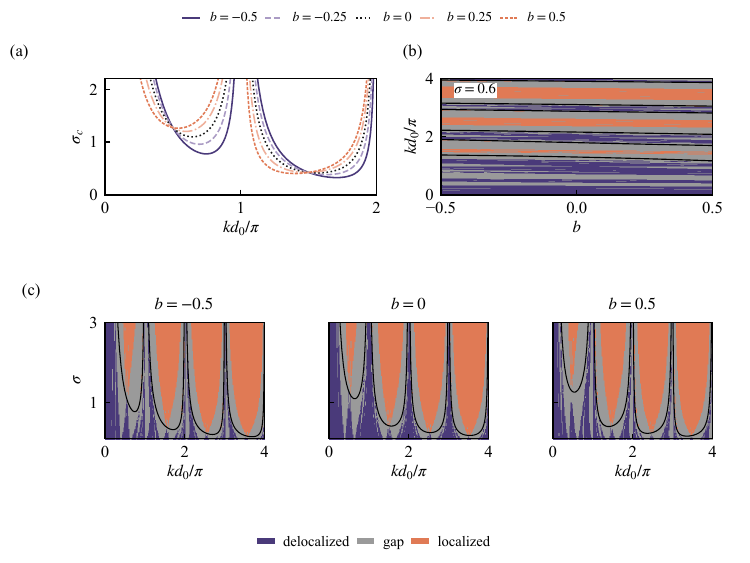}
\caption{Pulse-shape dependence of the GAA mobility edge. (a) Exact critical pulse areas for different $b$, shown by colored curves. (b) Classification at $\sigma=0.6$ as $b$ varies. (c) Classifications for $b=-0.5$, $0$, and $0.5$. Purple, gray, and orange denote delocalized, gap, and localized states in the maps; black curves in the maps are the analytic edges.}
\label{fig:sm_gaa}
\end{figure}

The leading shape deformation follows from
\begin{equation}
 \frac{\cos X}{1-b\cos X}
 =\cos X+\frac b2\cos 2X+\frac b2+O(b^2),
 \label{eq:sm_gaa_harmonic}
\end{equation}
so $b$ adds a second harmonic and a constant offset while leaving the free
propagation phase unchanged.  At $\sigma=0.6$, for $|b|\le0.5$ and
$0<kd_0<4\pi$, the displacement from the $b=0$ edge remains within about
one third of the adjacent delocalized-window width (Fig.~\ref{fig:sm_gaa}).
All edge curves $\sigma_c(kd_0,b)$ coincide at
$kd_0=(n+\tfrac12)\pi$, where $\cos(kd_0)=0$. Moreover,
\begin{equation}
 \frac{\partial\sigma_c}{\partial b}
 =-\frac{2\cos(kd_0)\operatorname{sgn}[\sin(kd_0)]}
 {kd_0|\sin(kd_0)|}.
 \label{eq:sm_gaa_direction}
\end{equation}
In each lobe $n\pi<kd_0<(n+1)\pi$, this derivative is negative on the
lower-wave-number half and positive on the upper half. Thus positive $b$
lowers and raises the critical area on those respective sides, as shown in
Fig.~\ref{fig:sm_gaa}(a).
The deformation $g=\cos X+\mu\cos2X$, where $\mu$ here is the second-harmonic amplitude, shifts the edge in the same direction as $b=2\mu$ at small $\mu$.

\section{Frequency comb}
\label{sec:sm_ipr}
\subsection{Continuous-time frequency comb}
\label{sec:sm_comb}
Suppose an analytic change of basis reduces the layer cocycle to a constant rotation:
\begin{equation}
 A(X)=B(X+2\pi\alpha)R(2\pi\rho)B(X)^{-1},\qquad
 \mathbf v_m=B(X_m)R(2\pi\rho m)B(X_0)^{-1}\mathbf v_0.
 \label{eq:sm_reduction}
\end{equation}
Here $\rho$ is the mean number of clockwise phase turns per layer (the rotation number), retaining the continuous lift of the angle. The matrices $B_n$ below are the Fourier coefficients of the reducing change of basis $B(X)$. Expanding
$B(X)=\sum_nB_ne^{inX}$ produces $n\alpha$, and the two rotation eigenmodes
produce $\pm\rho$. Inside layer $m$, with $s=(t-m\tau)/\tau\in[0,1)$,
\begin{equation}
 D(m\tau+s\tau)=D_m\cos[kd_0z(X_m)s]
       +\frac{(\dot D/\omega_0)_m}{z(X_m)}\sin[kd_0z(X_m)s].
 \label{eq:sm_continuous_field}
\end{equation}
The layer clock repeats with period $\tau$. Expanding its periodic dependence
after factoring out the rotation and modulation phases adds integer harmonics $j$,
so the continuous field has frequencies
\begin{equation}
 f\tau=\pm\rho+n\alpha+j,\qquad n,j\in\mathbb Z.
 \label{eq:sm_comb_rule}
\end{equation}
Here $f$ denotes temporal frequency, while $f(X)$ elsewhere denotes the modulation profile. The set of these integer-combination frequencies is the frequency module. For a gap, analytic stable and unstable directions reduce the cocycle to a
hyperbolic constant, allowing a double cover when the invariant line is not
orientable. The rotation number is locked to $2\rho\in\alpha\mathbb Z+\mathbb Z$;
this label accounts for the sign or cover, while the unstable solution acquires
the additional factor $e^{Lt/\tau}$. Its envelope-removed spectrum therefore
has the locked frequency module of Eq.~(\ref{eq:sm_comb_rule}).

Reducibility holds rigorously for analytic cocycles close to a constant rotation with Diophantine rotation number~\cite{Eliasson1992,HouYou2012}; at $r=0.6$ we test the comb numerically.

\subsection{Spectral definitions}
\begin{figure}[t]
\centering\includegraphics[width=\smmultifigwidth]{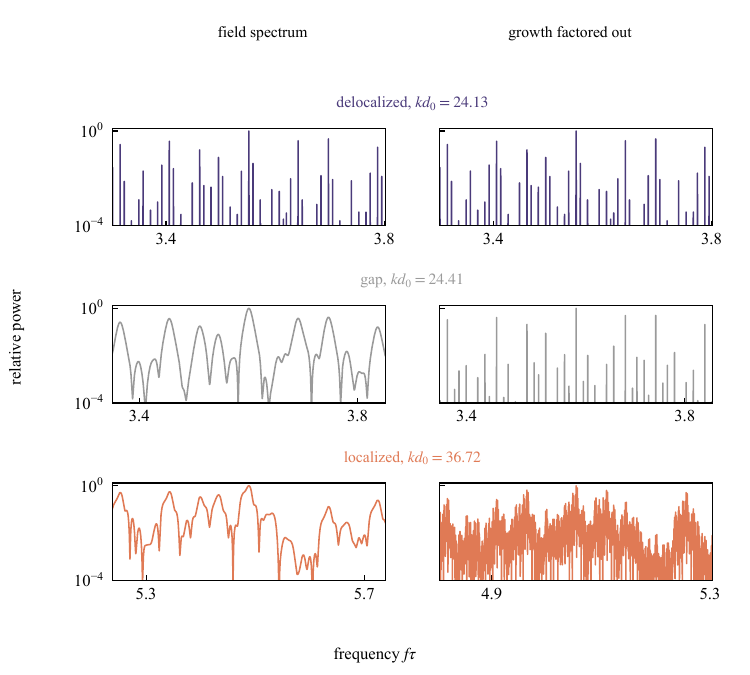}
\caption{Spectra of the field (left) and after factoring out the growth (right), for the delocalized, gap, and localized wave numbers of Fig.~3(a), at $r=0.6$. Each spectrum is normalized to its maximum and shown in a window around its strongest peak.}
\label{fig:sm_abc}
\end{figure}

We compare the power spectrum of the field $D(t)$ itself with the power spectrum after factoring out the overall growth $e^{Lt/\tau}$. For a growing field, the latter is the Laplace transform on $\operatorname{Re}s=L/\tau$, where $s$ is the complex Laplace frequency, evaluated through the Fourier transform of $e^{-Lt/\tau}D(t)$. This is the spectrum shown in Fig.~3(a).
For $n_s$ samples $u_\ell$ of the real field at interval $\Delta t$, $\ell$ labels the time sample and $j$ labels the Fourier frequency bin. The Hann window $W_\ell$ smoothly suppresses both ends of the finite time interval; the windowed power spectrum is
\begin{equation}
 p_j=\left|\sum_{\ell=0}^{n_s-1}W_\ell u_\ell e^{-2\pi i j\ell/n_s}\right|^2,\qquad
 W_\ell=\frac{1-\cos[2\pi\ell/(n_s-1)]}{2}.
 \label{eq:sm_power_spectrum}
\end{equation}
We use nonnegative frequencies, with spacing $\Delta(f\tau)=1/N$.
When the field grows, the spectrum of $D(t)$ is dominated by the last $\sim1/L$ layers, which broadens every line to a width of order $L/\pi$ in $f\tau$. Gap lines then merge, and a gap spectrum can no longer be told apart from a localized one. Factoring out the growth restores the lines of a gap (Fig.~\ref{fig:sm_abc}).

For the delocalized wave number $kd_0=24.13$, $\rho=4.170$. The 20 strongest peaks lie within half a frequency bin of Eq.~(\ref{eq:sm_comb_rule}), for both the delocalized and the gap wave number (Fig.~\ref{fig:sm_comb}). For a gap with small growth, $L=0.012$ at $kd_0=24.60$, the lines remain visible even without factoring out the growth (Fig.~\ref{fig:sm_smallgap}).

\subsection{Finite bandwidth and harmonic decay}
\begin{figure}[t]
\centering\includegraphics[width=\smmultifigwidth]{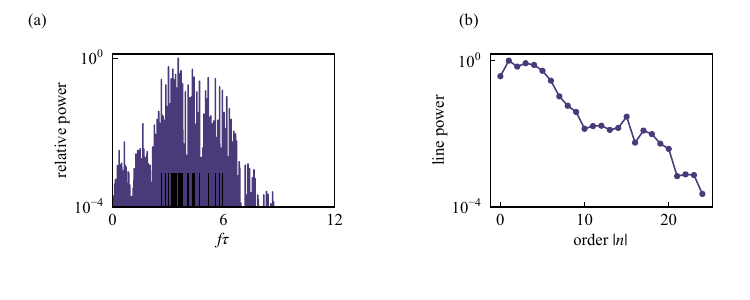}
\caption{(a) Spectrum after factoring out the growth at $kd_0=24.13$; black ticks mark predicted positions matched to the 20 strongest peaks using Eq.~(\ref{eq:sm_comb_rule}). (b) Resolved sideband powers at the same wave number. Both curves are calculated from the field and normalized to their respective maxima.}
\label{fig:sm_comb}
\label{fig:sm_sidebands}
\end{figure}

\begin{figure}[t]
\centering\includegraphics[width=\smfigwidth]{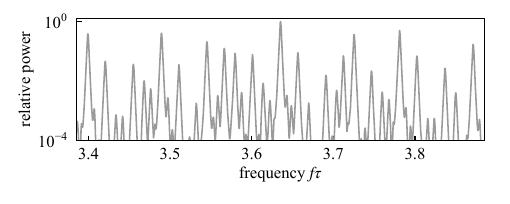}
\caption{Field spectrum without removing the growth for the small-growth gap at $kd_0=24.60$, $L=0.012$.}
\label{fig:sm_smallgap}
\end{figure}

\begin{figure}[t]
\centering\includegraphics[width=\smmultifigwidth]{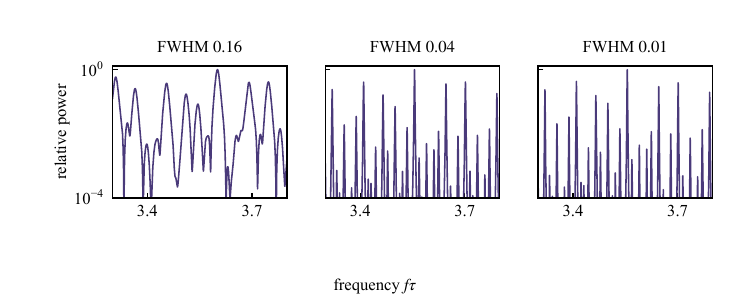}
\caption{Field spectra for Gaussian intensity FWHM 0.16, 0.04, and 0.01 (left to right), centered at $kd_0=24.13$.}
\label{fig:sm_bandwidth}
\end{figure}

The comb remains resolved for intensity full width at half maximum (FWHM), measured in $kd_0$, up to 0.16, the bandwidth of Fig.~3 (Fig.~\ref{fig:sm_bandwidth}).
If $B$ is analytic in $|\operatorname{Im}X|<h_*$, Cauchy's estimate gives
$\|B_n\|\leq C_he^{-h|n|}$ for each $h<h_*$, where $C_h$ bounds $B$ on that narrower strip. The asymptotic coefficient envelope determines the maximal analytic strip; individual observable lines can decay faster through projection or cancellation. A bounded analytic conjugacy to a rotation implies $L(h)=0$ on every narrower strip, so $h_*\leq h_{\rm sub}$. Fits to the resolved sidebands are compatible with this bound but do not determine $h_*$ quantitatively (Fig.~\ref{fig:sm_sidebands}).

\subsection{Phase classification and wave-number bands}
At $kd_0=36.72$, $L(0)=0.040$ and $L(h)$ rises with unit slope from $h=0$, so $\nu=1$.
The delocalized input band contains a minority of gap wave numbers (about one sixth), and the localized band is almost entirely localized (about 98 percent).
Figure~3(a) shows the localized wave number 36.72, and the localized band of Fig.~3(b)--(d) is centered at 104.41.

\section{Complexification and acceleration}
\label{sec:sm_avila}

Let $\Omega=2\pi\alpha$ and $|h|<h_{\rm sing}$. The complexified Lyapunov
exponent is
\begin{equation}
 L(h)=\lim_{N\to\infty}\frac1N\int_0^{2\pi}
 \ln\|A(X+(N-1)\Omega+ih)\cdots A(X+ih)\|\,
 \frac{\dd X}{2\pi}.
 \label{eq:sm_L}
\end{equation}
Avila's global theory makes $L(h)$ even, convex, and piecewise affine with
integer slopes [Theorem~5 of Ref.~\cite{Avila2015}], so $\nu=\partial_h^+L(0)$ is a nonnegative integer. The
complexified permittivity is
\begin{equation}
 \epsilon(X+ih)=\epsilon_R[1+r\sin X\cosh h+ir\cos X\sinh h].
 \label{eq:sm_complex_eps}
\end{equation}
The imaginary part is a quadrature gain/loss probe, and $h\to0^+$ measures the
response of the growth rate. In the non-Hermitian photonic time quasicrystal of
Ref.~\cite{Chen2026NHPTQC} this imaginary part is physical and $\nu$ quantizes
the amplification; here $h\to0^+$ serves as a probe.

For the analytic family, the spectrum is the set of wave numbers where
$(\alpha,A)$ is not uniformly hyperbolic; its complement is the set of momentum
gaps, following the standard cocycle/spectral correspondence~\cite{Avila2015,JohnsonMoser1982}.
We use
\begin{itemize}
\item delocalized: $L(h)=0$ on $|h|<h_{\rm sub}$ with $h_{\rm sub}>0$;
\item gap: $L(0)>0$ and $\nu=0$;
\item localized: $L(0)>0$ and $\nu\geq1$.
\end{itemize}
In Avila's terminology the localized phase is the supercritical regime.
For an analytic $SL(2,\mathbb C)$ cocycle at irrational frequency and $L(0)>0$, $\nu=0$ is equivalent to uniform hyperbolicity [Theorem~6 of Ref.~\cite{Avila2015}],
because a real cocycle with zero acceleration is regular near $h=0$. An invariant
unstable direction $E^u(X)$ has an angle $\psi(X)$ of degree $\ell$ in
$\mathbb{RP}^1$, the space of unoriented real lines; $\ell\in\mathbb Z$ counts the winding of that line. The conjugacy
\begin{equation}
 T(X)=R[\psi(X+\Omega)]^{-1}A(X)R[\psi(X)]
 =\begin{pmatrix}\mu(X)&c(X)\\0&\mu(X)^{-1}\end{pmatrix}
 \label{eq:sm_gap_triangular}
\end{equation}
has diagonal stretch $\mu(X)$ and off-diagonal shear $c(X)$, both scalar matrix entries. Its invariant direction locks the rotation number to
\begin{equation}
 2\rho\in \ell\alpha+\mathbb Z\subset\alpha\mathbb Z+\mathbb Z .
 \label{eq:sm_gap_label}
\end{equation}
This is the gap-labeling/rotation-number mechanism for almost-periodic
transfer problems [Ref.~\cite{AvilaJitomirskaya2010}, Theorem 2.2,
see also Ref.~\cite{JohnsonMoser1982}].
Thus the spectrum at a gap wave number has a locked frequency label, while a localized field has
positive acceleration and no such gap locking.

The analytic-strip and Fourier-coefficient relation is derived in
Sec.~\ref{sec:sm_comb}. 

\section{Mobility edge in momentum}
\label{sec:sm_edge}

\begin{figure}[t]
\centering\includegraphics[width=\smmultifigwidth]{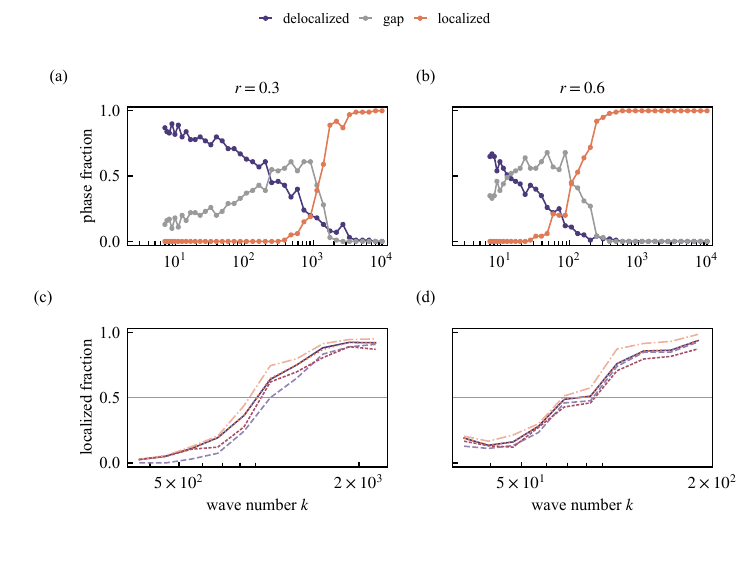}
\caption{Phase fractions and robustness of the mobility edge. Top: delocalized, gap, and localized fractions at $r=0.3$ and $0.6$. Bottom: localized share with gaps excluded for the default classifier (solid), halved $h$ step (dashed), doubled $N$ (dotted), and growth thresholds $0.2L_\infty$ (dash-dotted) and $0.4L_\infty$ (short-dashed).}
\label{fig:sm_fractions}
\label{fig:sm_robust}
\end{figure}

The localized share among delocalized and localized states, with gaps
excluded, and its logistic midpoint define the mobility edge.
Figure~\ref{fig:sm_fractions} shows the three sampled phase fractions: the numbers of delocalized, gap, and localized wave numbers divided by the total number in the same wave-number window. Thus the three fractions sum to one. The localized fraction in Fig.~2(f) and the lower row of Fig.~\ref{fig:sm_fractions} instead divides the localized count by the combined delocalized and localized count, excluding gaps.
The classifier uses
\begin{equation}
 s_\nu=\frac{L(0.004)-L(0.002)}{0.002},
 \label{eq:sm_slope}
\end{equation}
which estimates the first affine slope of $L(h)$. With the high-wave-number random-phase estimate
\begin{equation}
 L_\infty(r)=\frac1{2\pi}\int_0^{2\pi}\ln\cosh\eta(X)\,\dd X,\qquad
 \eta(X)=\frac14\ln\frac{\epsilon(X+\Omega)}{\epsilon(X)},
 \label{eq:sm_linf}
\end{equation}
states with $L(0)<0.3L_\infty$ are delocalized; states with
$L(0)\geq0.3L_\infty$ and $s_\nu\geq0.6$ are localized; the remainder are gaps.

A random-phase approximation at high wave number treats
$kd_0z(X_m)$ as uniformly distributed independently of the incoming
projective direction. For the random-phase
product $H(\eta_m)R(\theta_m)$, with $\theta_m=kd_0z(X_m)$ and incoming direction angle $\vartheta$, the rotation randomizes the direction before the boost; averaging this
uniform incoming direction gives
\begin{equation}
 L_\infty=\left\langle\frac1{2\pi}\int_0^{2\pi}
 \frac12\ln(e^{-2\eta}\cos^2\vartheta+e^{2\eta}\sin^2\vartheta)\,\dd\vartheta
 \right\rangle_X
 =\langle\ln\cosh\eta\rangle_X .
 \label{eq:sm_linf_random}
\end{equation}
Gap states are removed before forming
\begin{equation}
 F_{\rm loc}(kd_0,r)=\frac{f_{\rm loc}}{f_{\rm del}+f_{\rm loc}},\qquad
 F_{\rm fit}=\frac{1}{1+\exp[-(\ln(kd_0)-\ln[(kd_0)_{1/2}])/w]} .
 \label{eq:sm_Q}
\end{equation}
Here $f_{\rm del}$ and $f_{\rm loc}$ are the delocalized and localized fractions
in a wave-number window, and $F_{\rm fit}$ is fitted to those window values. The fitted midpoint $(kd_0)_{1/2}$ is the wave number at which $F_{\rm fit}=1/2$.
The midpoint moves from $(kd_0)_{1/2}\simeq70$ at $r=0.6$ to
$1.2\times10^4$ at $r=0.1$, and for $r\leq0.2$ it grows as $r^{-2.2}$.
Here $w$ is the transition width in $\ln(kd_0)$.
The golden mean $\alpha=(\sqrt5-1)/2$ and silver mean $\alpha=\sqrt2-1$ give midpoints that agree within the fitted transition width at $r=0.2$, $0.3$, and $0.6$. Sensitivity to halving $h$, doubling $N$, and replacing $0.3L_\infty$ by $0.2L_\infty$ or $0.4L_\infty$ is shown in Fig.~\ref{fig:sm_robust}.

\section{Normal form and the weak-modulation bound}
\label{sec:sm_normal}

This section proves the weak-modulation bound in the main text. We state the result for a
general modulation $\epsilon(X)=\epsilon_R[1+rf(X)]$ and then specialize.

\paragraph{Assumptions.}
(A1) $f$ is real-analytic and $2\pi$-periodic, and extends holomorphically to
the strip $|\operatorname{Im}X|<h_f$; fix $h_0\in(0,h_f)$ and let
$M=\sup_{|\operatorname{Im}X|\leq h_0}|f(X)|$.
(A2) $0<r\leq r_*:=1/(2M)$, so that $|rf(X)|\leq\tfrac12$ on the closed strip
$|\operatorname{Im}X|\leq h_0$.
(A3) $\alpha$ is Diophantine: there are $\gamma>0$ and $\beta\geq1$ with
\begin{equation}
 \|p\alpha\|_{\mathbb R/\mathbb Z}\geq\frac{\gamma}{|p|^{\beta}}\qquad(p\neq0).
 \label{eq:sm_DC}
\end{equation}
The golden mean satisfies Eq.~(\ref{eq:sm_DC}) with $\beta=1$ and
$\gamma=1/3$, since it is badly approximable. With $\Omega=2\pi\alpha$,
Eq.~(\ref{eq:sm_DC}) gives
$|e^{ip\Omega}-1|=2|\sin(\pi p\alpha)|\geq4\|p\alpha\|\geq4\gamma|p|^{-\beta}$.

Under (A1)--(A2), the functions
\begin{equation}
 z(X)=[1+rf(X)]^{-1/2},\qquad
 \eta(X)=\frac14\ln\frac{1+rf(X+\Omega)}{1+rf(X)}
 \label{eq:sm_z_eta}
\end{equation}
are holomorphic on $|\operatorname{Im}X|<h_0$ (principal branches), real on the
real axis, and satisfy
\begin{equation}
 |z(X)-1|\leq C_1r,\qquad |\eta(X)|\leq C_2r
 \qquad(|\operatorname{Im}X|<h_0),
 \label{eq:sm_z_eta_bounds}
\end{equation}
with $C_1,C_2$ depending only on $M$. Here $z(X_m)=d_m/d_0$, so the phase
accumulated in layer $m$ is $kd_m=kd_0z(X_m)$, and $\eta(X_m)=\eta_m$.

\begin{proposition}
Under (A1)--(A3) there are constants $c_1,c_2>0$ and $r_0\in(0,r_*]$,
depending only on $\alpha$ and $f$, such that for $0<r<r_0$ every wave number
with
\begin{equation}
 kd_0<\frac{c_1}{r}\ln\frac{c_2}{r}
 \label{eq:sm_onset_prop}
\end{equation}
has acceleration $\nu=0$. Such a wave number is delocalized or lies in a gap,
and the localized phase ($\nu\geq1$) requires
$kd_0\geq(c_1/r)\ln(c_2/r)$.
\end{proposition}

The proof has five steps. Throughout, two cocycles $(\alpha,A)$ and
$(\alpha,A')$ are conjugate if $A'(X)=B(X+\Omega)A(X)B(X)^{-1}$ with $B$
holomorphic on a strip $|\operatorname{Im}X|<h$ and $B,B^{-1}$ bounded there.
Products then satisfy
$A'_N(X)=B(X+N\Omega)A_N(X)B(X)^{-1}$, so $L(h')$ and the acceleration
$\nu$ are unchanged for $|h'|<h$.

\paragraph{Step 1: normal form.}
With $S(X)=\operatorname{diag}[\sqrt{z(X)},1/\sqrt{z(X)}]$ and
$R(\theta)$ as in Eq.~(\ref{eq:sm_layer}), the layer matrix is
$A(X)=S(X)^{-1}R[kd_0z(X)]S(X)$. Conjugating by $S$,
\begin{equation}
 \widetilde A(X)=S(X+\Omega)A(X)S(X)^{-1}=H[\eta(X)]\,R[kd_0z(X)],
 \qquad H(\eta)=\operatorname{diag}(e^{-\eta},e^{\eta}),
 \label{eq:sm_normal}
\end{equation}
because $S(X+\Omega)S(X)^{-1}=\operatorname{diag}
\{[z(X+\Omega)/z(X)]^{1/2},[z(X)/z(X+\Omega)]^{1/2}\}=H[\eta(X)]$.
This is the normal form in the main text. $S$ and $S^{-1}$ are bounded on the strip by
Eq.~(\ref{eq:sm_z_eta_bounds}).

\paragraph{Step 2: cohomological equation.}
Let $\bar z=(2\pi)^{-1}\int_0^{2\pi}z\,\dd X$ and $z_p$ be the Fourier
coefficients of $z$. Since $|z-1|\leq C_1r$ on the strip, Cauchy's estimate
gives $|z_p|\leq C_1r\,e^{-h_0|p|}$ for $p\neq0$. The equation
\begin{equation}
 \Psi(X+\Omega)-\Psi(X)=z(X)-\bar z
 \label{eq:sm_cohomology}
\end{equation}
is solved by $\Psi_0=0$ and $\Psi_p=z_p/(e^{ip\Omega}-1)$ for $p\neq0$. On the
narrower strip $|\operatorname{Im}X|\leq h_1:=h_0/2$,
\begin{equation}
 |\Psi(X)|\leq\sum_{p\neq0}\frac{C_1r\,|p|^{\beta}}{4\gamma}
 e^{-h_0|p|/2}\leq C_\alpha r,\qquad
 C_\alpha=\frac{C_1}{2\gamma}\sum_{p\geq1}p^{\beta}e^{-h_0p/2}.
 \label{eq:sm_cohomology_bound}
\end{equation}
$\Psi$ is real on the real axis because $z$ is.

\paragraph{Step 3: removal of the phase variation.}
Rotations commute, so
\begin{equation*}
 R[kd_0z(X)]=R[kd_0\Psi(X+\Omega)]R(kd_0\bar z)R[-kd_0\Psi(X)].
\end{equation*}
Conjugating by $C(X)=R[kd_0\Psi(X)]$ gives
\begin{equation}
 \widehat A(X)=C(X+\Omega)^{-1}\widetilde A(X)C(X)
 =R(-\vartheta)\,H[\eta(X)]\,R(\vartheta)\;R(\theta_0),\qquad
 \vartheta=kd_0\Psi(X+\Omega),\quad\theta_0=kd_0\bar z .
 \label{eq:sm_phase_removed}
\end{equation}
For complex $X$, $C$ is not unitary, but $\|C\|,\|C^{-1}\|\leq
e^{kd_0C_\alpha r}$ on $|\operatorname{Im}X|\leq h_1$, which is finite for
fixed $k$ and $r$. The conjugacy therefore preserves $L(h)$ and $\nu$.

\paragraph{Step 4: distance from a constant rotation.}
Writing out the first factor,
\begin{equation}
 R(-\vartheta)H(\eta)R(\vartheta)
 =\cosh\eta\,\mathbb{I}-\sinh\eta
 \begin{pmatrix}\cos2\vartheta&\sin2\vartheta\\
 \sin2\vartheta&-\cos2\vartheta\end{pmatrix}.
 \label{eq:sm_rotated_H}
\end{equation}
On $|\operatorname{Im}X|\leq h_1$, $|\eta|\leq C_2r$ and
$|\operatorname{Im}\vartheta|\leq kd_0C_\alpha r$, so
$|\cos2\vartheta|,|\sin2\vartheta|\leq e^{2kd_0C_\alpha r}$ and
\begin{equation}
 \sup_{|\operatorname{Im}X|\leq h_1}
 \bigl\|\widehat A(X)-R(\theta_0)\bigr\|
 \leq C_3\,r\,e^{2C_\alpha kd_0r},
 \label{eq:sm_bound}
\end{equation}
with $C_3$ depending only on $M$ (for $r\leq r_*$). The interface factor is
$O(r)$, while the phase variation enters only through $2\vartheta=O(kd_0r)$.
This is the perturbation of size $r\,e^{O(kd_0r)}$ in the main text.

\paragraph{Step 5: a uniform neighborhood of the constant rotations.}
Let $\mathcal B_{h_1}$ be the Banach space of $2\times2$ matrix functions,
$2\pi$-periodic and holomorphic on $|\operatorname{Im}X|<h_1$, continuous up to
the boundary, with the sup norm, and fix $h_2=h_1/2$. For a cocycle
$A'\in\mathcal B_{h_1}$ with values in $SL(2,\mathbb C)$:
\begin{enumerate}
\item $L_{A'}(h)$ is convex in $h$, and its one-sided slopes are quantized
 (the acceleration is an integer for irrational frequency) [Theorem~5 of Ref.~\cite{Avila2015}].
 For real cocycles $L_{A'}(h)$ is even, so $\nu=\partial_h^+L_{A'}(0)\geq0$, and convexity gives
 $\nu\leq[L_{A'}(h_2)-L_{A'}(0)]/h_2$.
\item $L_{A'}(0)$ and $L_{A'}(h_2)$ depend continuously on $A'\in\mathcal B_{h_1}$
 for irrational $\alpha$ [Corollary~2 of Ref.~\cite{JKS2009}]; $L_{A'}(h_2)$ is the Lyapunov exponent
 of $X\mapsto A'(X+ih_2)$, which is holomorphic on
 $|\operatorname{Im}X|<h_1-h_2$ and depends continuously on $A'$.
\item A constant rotation has $L(h)=0$ for all $h$.
\end{enumerate}
The set $U=\{A':\ L_{A'}(h_2)-L_{A'}(0)<h_2/2\}$ is open by item 2 and
contains every constant rotation by item 3. By item 1 and quantization, every
real $A'\in U$ has $\nu=0$. The constant rotations
$\mathcal K=\{R(\theta_0):\theta_0\in[0,2\pi]\}$ form a compact subset of $U$,
so their distance to the complement of $U$ is positive. Hence there is
$\varepsilon_0>0$, depending only on $\alpha$ and $h_1$, such that
\begin{equation}
 \sup_{|\operatorname{Im}X|\leq h_1}\|A'(X)-R(\theta_0)\|<\varepsilon_0
 \ \text{for some }\theta_0
 \quad\Longrightarrow\quad\nu(A')=0 .
 \label{eq:sm_uniform}
\end{equation}
Uniformity in $\theta_0$ matters because $\theta_0=kd_0\bar z$ changes with
$k$.

\paragraph{Conclusion of the proof.}
By Eqs.~(\ref{eq:sm_bound}) and (\ref{eq:sm_uniform}), $\nu(\widehat A)=0$
whenever $C_3re^{2C_\alpha kd_0r}<\varepsilon_0$, that is, whenever
\begin{equation}
 kd_0<\frac{c_1}{r}\ln\frac{c_2}{r},\qquad
 c_1=\frac1{2C_\alpha},\qquad c_2=\frac{\varepsilon_0}{C_3}.
 \label{eq:sm_onset}
\end{equation}
Steps 1 and 3 preserve $\nu$, so $\nu(A)=0$. The constants depend on $\alpha$
through $\gamma,\beta,\varepsilon_0$ and on $f$ through $M,h_0$; take
$r_0=\min(r_*,c_2)$ so that the logarithm is positive. \hfill$\square$

\paragraph{Equal-phase layers.}
For layer durations $\tau_m=\tau\sqrt{\epsilon_m/\epsilon_R}$ every layer
accumulates the phase $kd_0$, so Step 1 gives
$\widetilde A(X)=H[\eta(X)]R(kd_0)$ with no phase variation, and
$\|\widetilde A-R(kd_0)\|\leq C_4r$ on the strip, uniformly in $k$. By
Eq.~(\ref{eq:sm_uniform}), $\nu=0$ at every wave number once
$C_4r<\varepsilon_0$. This is the statement used for the equal-phase
protocol in the main text.

\paragraph{Remarks.}
(i) The bound excludes positive acceleration below the onset, while allowing momentum gaps. That the
localized phase indeed appears above the edge is established numerically
(Sec.~\ref{sec:sm_edge}).
(ii) The numerically determined edge, $kd_0\propto r^{-2.2}$, lies above the
bound, as it must.
(iii) A first-harmonic estimate locates the edge more closely than the
rigorous bound. With $g=\ln(\epsilon/\epsilon_R)$, one has
$\eta_p=\tfrac14g_p(e^{ip\Omega}-1)$ and $\Psi_p=z_p/(e^{ip\Omega}-1)$, so in
the product $2kd_0\Psi_p\eta_p$ the small denominator cancels,
\begin{equation}
 2kd_0|\Psi_p||\eta_p|=\frac{kd_0}2|z_p||g_p| .
 \label{eq:sm_harmonic}
\end{equation}
For $f=\sin$, the leading-order first-harmonic amplitudes are
$\widehat\eta=\tfrac r2|\sin\pi\alpha|$ and
$\zeta=kd_0r/(2|\sin\pi\alpha|)$, with $\zeta\widehat\eta=kd_0r^2/4+O(kd_0r^3)$.
Setting this product to a constant suggests $kd_0\propto r^{-2}$ and an edge
nearly independent of $\alpha$, consistent with the fitted $r^{-2.2}$ and with
the close agreement between golden and silver means. This estimate is
heuristic and is not used in the proof.

\section{Two-band propagation}
\label{sec:sm_twoband}

For a complex solution of Eq.~(\ref{eq:sm_wave}), the Wronskian
$W=D_k^*\dot D_k-D_k\dot D_k^*$ is constant within a layer and continuous at
interfaces. Hence $\operatorname{Im}(D_k^*\dot D_k)$ is conserved. With
\begin{equation}
 a_\pm=\frac12\left(D_k\pm\frac{i\dot D_k}{\omega_0}\right),
 \label{eq:sm_ports}
\end{equation}
\begin{equation}
 |a_+|^2-|a_-|^2
 =-\frac1{\omega_0}\operatorname{Im}(D_k^*\dot D_k).
 \label{eq:sm_wronskian}
\end{equation}
Thus $P_f-P_b$ is conserved mode by mode. With
$P_{f,b}=\sum_k|a_\pm(k)|^2$ denoting forward and backward band powers in the reference normalization and $P(t)=P_f(t)+P_b(t)$ their sum, a forward input with initial power $P(0)$ gives the exact identity
\begin{equation}
 \frac{P_f-P_b}{P_f+P_b}=\frac{P(0)}{P(t)}.
 \label{eq:sm_vg}
\end{equation}
After the modulation is stopped and the pulse propagates in the reference
medium, its total intensity centroid has velocity
$v_g=v_R[P_f-P_b]/[P_f+P_b]$; this follows by integrating the
reference-medium intensity continuity equation. The transfer-matrix gain of 68~dB gives $v_g/v_R\approx1.5\times10^{-7}$. The Yee-FDTD calculation gives 69~dB for the localized band and 2.5~dB for the delocalized band.

During modulation, the intensity used for Fig.~3(c) is
$I=(|D|^2+|b|^2)/2$, where $b=B/(\mu v_R)$ is the scaled magnetic induction and
$a(t)=\epsilon_R/\epsilon(t)=z(t)^2$. With space and time measured in $d_0$ and $\tau$, respectively, the intensity flux $J$ and modulation source $\mathcal S$ obey
\begin{equation}
 \partial_t I+\partial_x J=\mathcal S,\qquad
 J=\frac{1+a}{2}\operatorname{Re}(D^*b),\qquad
 \mathcal S=\frac{a-1}{2}\operatorname{Re}(D^*\partial_xb-b^*\partial_xD).
 \label{eq:sm_reference_continuity}
\end{equation}
The intensity centroid is $\bar x=\int xI\,\dd x/\int I\,\dd x$. For a packet with negligible boundary terms its first moment obeys
\begin{equation}
 \frac{\dd\bar x}{\dd t}
 =\frac{\int J\,\dd x}{\int I\,\dd x}
  +\frac{\int(x-\bar x)\mathcal S\,\dd x}{\int I\,\dd x}.
 \label{eq:sm_centroid_source}
\end{equation}
Restoring units multiplies the right-hand side by $v_R$. At $a=1$ the source
vanishes and the identity gives the centroid velocity.
For the modulated pulse, the source first moment and the factor $(1+a)/2$
must be retained. Figure~\ref{fig:sm_centroid_comparison} compares the
central-difference velocity from Fig.~3(c) with the power ratio.
During modulation the centroid velocity oscillates about this ratio (Fig.~\ref{fig:sm_centroid_comparison}).

\subsection{Yee discretization}
\begin{figure}[t]
\centering\includegraphics[width=\smmultifigwidth]{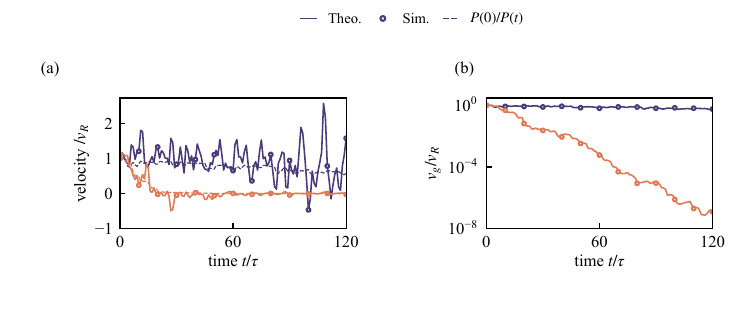}
\caption{(a) Centroid velocity during modulation for the two bands of Fig.~3; dashed curves show $P(0)/P(t)$. (b) Group velocity after the modulation is switched off at each indicated time. Purple and orange denote the delocalized and localized bands. Theo. (lines): transfer matrices; Sim. (circles): staggered-grid Yee-FDTD calculation.}
\label{fig:sm_centroid_comparison}
\label{fig:sm_velocity}
\end{figure}

The finite-difference time-domain (FDTD) method updates the two Maxwell fields on staggered space and time grids.
In units $x/d_0$ and $t/\tau$, Maxwell's equations read $\partial_tD=-\partial_xb$ and $\partial_tb=-a(t)\partial_xD$, with $a=z^2$.
A Yee grid places $D$ at integer points and times and $b$ at half-grid points and half-times:
\begin{equation}
 \begin{split}
 b^{n+1/2}_{i+1/2}&=b^{n-1/2}_{i+1/2}-a^nS(D^n_{i+1}-D^n_i),\\
 D^{n+1}_i&=D^n_i-S(b^{n+1/2}_{i+1/2}-b^{n+1/2}_{i-1/2}),
 \end{split}
 \label{eq:sm_fdtd_update}
\end{equation}
where $i$ and $n$ index spatial cells and time steps, $\Delta x$ and $\Delta t$ are their dimensionless spacings, $a^n=a(n\Delta t)$, and $S=\Delta t/\Delta x$. With $a_{\max}=\max_t a(t)$, the maximum-speed Courant number is $\sqrt{a_{\max}}S$. Stability requires $\sqrt{a_{\max}}\,\Delta t/\Delta x\leq1$; the phase error is second order in $\Delta x$. Figure~\ref{fig:sm_velocity} compares the two propagation calculations and the group velocity after switching off the modulation.

\section{Protocol dependence}
\label{sec:sm_protocol}

\textit{Equal propagation phase.}---Layers with
$\tau_m=\tau\sqrt{\epsilon_m/\epsilon_R}$ accumulate the same phase
$kd_0$. The interface factors remain, while the wave number
enters only through $kd_0\bmod2\pi$. Here $\epsilon_{\max}$ and $\epsilon_{\min}$ are the largest and smallest layer permittivities. The conjugacy of Eq.~(\ref{eq:sm_normal})
gives $H[\eta(X)]R(kd_0)$ with $\|H(\eta)-\mathbb{I}\|\leq Cr$, where $C$ depends only on the modulation profile and analytic strip, uniformly in
$kd_0$, so the perturbative argument of Sec.~\ref{sec:sm_normal} gives $\nu=0$ at every wave
number for $r<\varepsilon_0/C$. In the finite scan, equal-phase samples show no
localized state up to $\epsilon_{\max}/\epsilon_{\min}\simeq2\times10^3$,
whereas equal-duration layers develop a localized fraction. 
\begin{figure}[t]
\centering\includegraphics[width=\smfigwidth]{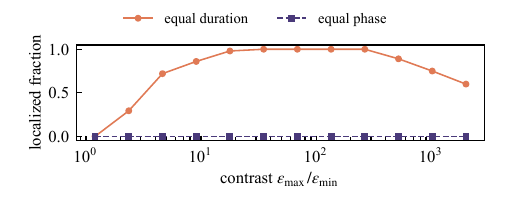}
\caption{Localized fraction versus permittivity contrast for equal-duration and equal-phase layers. The fraction counts localized wave-number--initial-phase pairs among all sampled pairs, with gaps included in the denominator. No equal-phase sample has $\nu\geq1$.}
\label{fig:sm_equalphase}
\end{figure}

\textit{Random order.}---For $0<r<1$ and $k>0$, drawing each $X_m$ independently and uniformly
gives positive growth. The distribution has bounded matrix norms and thus
the required finite logarithmic moment. Each $A(X)$ is elliptic, conjugate by
$S(X)$ to the rotation $R[kd_0 z(X)]$, and preserves the quadratic form
$(d_m/d_0)D^2+(d_0/d_m)(\dot D/\omega_0)^2$, which differs between layers of different
permittivity. For a nontrivial elliptic rotation this positive quadratic
form is unique up to scale; choose two such layers with different
permittivities. Their forms cannot be proportional. A compact subgroup of $SL(2,\mathbb R)$ preserves a single form,
so the group generated by the layers is noncompact. Because $kd_0 z(X)$ ranges over
an interval, the family contains rotations by irrational multiples of $\pi$,
which leave no finite set of lines invariant, so the group is strongly
irreducible. Furstenberg's theorem then gives $L>0$~\cite{Furstenberg1963}.
For the impulsive matrices $R(kd_0)J_m$ with $\sigma kd_0\sin(kd_0)\neq0$,
the group contains the shears $J_m^{-1}J_{m'}$, which make it noncompact and
fix only the line $(0,1)^{\mathsf T}$, a line that $R(kd_0)$ moves; the
theorem again gives $L>0$.

\begin{figure}[t]
\centering\includegraphics[width=\smmultifigwidth]{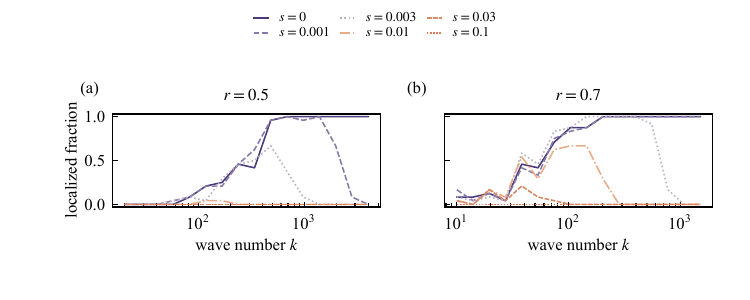}
\caption{Localized fraction for smoothstep switching at $r=0.5$ (a) and $0.7$ (b), for several switching fractions $s$. Here the localized fraction is the localized count divided by all wave numbers in each window, with gaps included in the denominator.}
\label{fig:sm_switching}
\end{figure}

\textit{Finite switching time.}---A smooth switch reduces interface mixing.
For a ramp with profile $3\xi^2-2\xi^3$, where $\xi=(t-t_{\rm start})/t_r\in[0,1]$ measures time from the ramp start $t_{\rm start}$ and $t_r=s\tau$ is its duration, first-order perturbation
theory gives
\begin{equation}
 R\simeq\frac{\Delta}{2}I(\chi),\qquad
 I(\chi)=\int_0^1 6\xi(1-\xi)e^{i\chi\xi}\,\dd\xi,\qquad \chi=2kd_0 z(X)s,
 \label{eq:sm_switch_reflection}
\end{equation}
where $\Delta=-2\eta$ is the first-order fractional frequency change across the interface, $s=t_r/\tau$ is the switching-time fraction,
and the approximation is first order in $|\Delta|\ll1$.
In an instantaneous forward/backward basis the off-diagonal coupling is
$\dot\omega/(2\omega)$; integrating it against the unperturbed phase
$e^{2i\omega t}$ gives the integral above, with the sign fixed by the
interface convention in Eq.~(\ref{eq:sm_fresnel_coeff}). Terms from varying
the phase during the switch are of second order in the small contrast.
Twice integrating by parts, using $6\xi(1-\xi)=0$ at both endpoints,
gives
\begin{equation}
 I(\chi)=6\left[-\frac{1+e^{i\chi}}{\chi^2}
  -\frac{2i(e^{i\chi}-1)}{\chi^3}\right].
 \label{eq:sm_switch_integral}
\end{equation}
Thus $I(0)=1$. The exact expression gives the rigorous bound
$|I(\chi)|\leq12/\chi^2+24/|\chi|^3$ for $\chi\ne0$ and
$I(\chi)=O(\chi^{-2})$ as $|\chi|\to\infty$. The ramp suppresses
the effective interface rapidity at high $kd_0  s$; the finite-switching scan shows that the localized phase is confined below a wave number that decreases as $s$ grows (Fig.~\ref{fig:sm_switching}).

\section{Numerical conventions and reproduction}
\label{sec:sm_params}
Growth rates are obtained by renormalizing the two-component state after each layer and accumulating the logarithms of the removed norms. Initial phases are averaged. The impulsive panels use the exact shear map and the layered panels use Eq.~(\ref{eq:sm_layer}).

\paragraph{Numerical parameters.}
The continuous spectra use $r=0.6$, $\phi=0$, $N=32768$ layers, 128 samples per layer, and a Hann window. The removed growth rate is the least-squares slope of the accumulated log norm. Comb positions use a rotation number evaluated over $2^{20}$ layers and harmonics $|n|\leq12$. Gaussian-band spectra use 64 samples per layer and 257 wave numbers over four intensity standard deviations. The $L(h)$ checks use four initial phases, up to $N=10^5$ layers, and $h=0,0.002,0.005,0.01$; the sideband strip-width scan uses $N=65536$. The input-band classification uses spacing $0.002$, four phases, $N=30000$, growth threshold $10^{-3}$, and slope threshold $0.6$.
For the mobility-edge scans, each width-10 window contains 100 wave numbers at spacing 0.1, averaged over eight phases $\phi_j=j/8+0.013$ modulo one. The classifier uses $h=0,0.002,0.004$, growth threshold $0.3L_\infty$, and slope threshold 0.6. Layer counts range from $N=6000$ at $r\geq0.45$ to $N\simeq1.2\times10^5$ at $r=0.1$. The golden-mean map has 40 logarithmically spaced windows; the small-$r$ and silver-mean scans use ten. Logistic fits minimize unweighted residuals in logarithmic wave number, with $0.02\leq w\leq3$, using windows with at least five spectral samples. The equal-phase scan uses contrast $(1+r)/(1-r)$ from 1.2 to $2.0\times10^3$, 512 propagation phases, eight initial phases, and $N=6000$; its equal-duration control uses 100 logarithmically spaced wave numbers in $kd_0\in[20,2000]$. Smoothstep scans use $s=0,0.001,0.003,0.01,0.03,0.1$, 24 samples per width-10 window, eight phases, $N=4000$, and the same classifier. The pulse of Fig.~3 has equal power in $kd_0\in[24.05,24.21]$ and $[104.33,104.49]$, an eighth-power spectral envelope (the amplitude is an exponential of minus the eighth power of the detuning normalized by the band half-width), a periodic interval of length $480d_0$, $r=0.6$, and $\phi=0$. Both calculations propagate 120 layers; the Yee-FDTD grid has $2^{20}$ cells and 3636 substeps per layer, with maximum-speed Courant number 0.95.

The scripts and data that generate all figures are available from the corresponding author.


\begin{thebibliography}{55}\makeatletter
\providecommand \@ifxundefined [1]{ \@ifx{#1\undefined}
}\providecommand \@ifnum [1]{ \ifnum #1\expandafter \@firstoftwo
 \else \expandafter \@secondoftwo
 \fi
}\providecommand \@ifx [1]{ \ifx #1\expandafter \@firstoftwo
 \else \expandafter \@secondoftwo
 \fi
}\providecommand \natexlab [1]{#1}\providecommand \enquote  [1]{``#1''}\providecommand \bibnamefont  [1]{#1}\providecommand \bibfnamefont [1]{#1}\providecommand \citenamefont [1]{#1}\providecommand \href@noop [0]{\@secondoftwo}\providecommand \href [0]{\begingroup \@sanitize@url \@href}\providecommand \@href[1]{\@@startlink{#1}\@@href}\providecommand \@@href[1]{\endgroup#1\@@endlink}\providecommand \@sanitize@url [0]{\catcode `\\12\catcode `\$12\catcode
  `\&12\catcode `\#12\catcode `\^12\catcode `\_12\catcode37=12\relax}\providecommand \@@startlink[1]{}\providecommand \@@endlink[0]{}\providecommand \url  [0]{\begingroup\@sanitize@url \@url }\providecommand \@url [1]{\endgroup\@href {#1}{\urlprefix }}\providecommand \urlprefix  [0]{URL }\providecommand \Eprint [0]{\href }\providecommand \doibase [0]{https://doi.org/}\providecommand \selectlanguage [0]{\@gobble}\providecommand \bibinfo  [0]{\@secondoftwo}\providecommand \bibfield  [0]{\@secondoftwo}\providecommand \translation [1]{[#1]}\providecommand \BibitemOpen [0]{}\providecommand \bibitemStop [0]{}\providecommand \bibitemNoStop [0]{.\EOS\space}\providecommand \EOS [0]{\spacefactor3000\relax}\providecommand \BibitemShut  [1]{\csname bibitem#1\endcsname}\let\auto@bib@innerbib\@empty
\bibitem [{\citenamefont {Lustig}\ \emph {et~al.}(2018)\citenamefont {Lustig},
  \citenamefont {Sharabi},\ and\ \citenamefont {Segev}}]{Lustig2018}  \BibitemOpen
  \bibfield  {author} {\bibinfo {author} {\bibfnamefont {E.}~\bibnamefont
  {Lustig}}, \bibinfo {author} {\bibfnamefont {Y.}~\bibnamefont {Sharabi}},\
  and\ \bibinfo {author} {\bibfnamefont {M.}~\bibnamefont {Segev}},\ }\bibfield
   {title} {\bibinfo {title} {{Topological aspects of photonic time
  crystals}},\ }\href {https://doi.org/10.1364/OPTICA.5.001390} {\bibfield
  {journal} {\bibinfo  {journal} {Optica}\ }\textbf {\bibinfo {volume} {5}},\
  \bibinfo {pages} {1390} (\bibinfo {year} {2018})}\BibitemShut {NoStop}\bibitem [{\citenamefont {Galiffi}\ \emph {et~al.}(2022)\citenamefont
  {Galiffi}, \citenamefont {Tirole}, \citenamefont {Yin}, \citenamefont {Li},
  \citenamefont {Vezzoli}, \citenamefont {Huidobro}, \citenamefont
  {Silveirinha}, \citenamefont {Sapienza}, \citenamefont {Al\`u},\ and\
  \citenamefont {Pendry}}]{Galiffi2022}  \BibitemOpen
  \bibfield  {author} {\bibinfo {author} {\bibfnamefont {E.}~\bibnamefont
  {Galiffi}}, \bibinfo {author} {\bibfnamefont {R.}~\bibnamefont {Tirole}},
  \bibinfo {author} {\bibfnamefont {S.}~\bibnamefont {Yin}}, \bibinfo {author}
  {\bibfnamefont {H.}~\bibnamefont {Li}}, \bibinfo {author} {\bibfnamefont
  {S.}~\bibnamefont {Vezzoli}}, \bibinfo {author} {\bibfnamefont {P.~A.}\
  \bibnamefont {Huidobro}}, \bibinfo {author} {\bibfnamefont {M.~G.}\
  \bibnamefont {Silveirinha}}, \bibinfo {author} {\bibfnamefont
  {R.}~\bibnamefont {Sapienza}}, \bibinfo {author} {\bibfnamefont
  {A.}~\bibnamefont {Al\`u}},\ and\ \bibinfo {author} {\bibfnamefont {J.~B.}\
  \bibnamefont {Pendry}},\ }\bibfield  {title} {\bibinfo {title} {Photonics of
  time-varying media},\ }\href {https://doi.org/10.1117/1.AP.4.1.014002}
  {\bibfield  {journal} {\bibinfo  {journal} {Adv. Photon.}\ }\textbf {\bibinfo
  {volume} {4}},\ \bibinfo {pages} {014002} (\bibinfo {year}
  {2022})}\BibitemShut {NoStop}\bibitem [{\citenamefont {Zurita-S\'anchez}\ \emph {et~al.}(2009)\citenamefont
  {Zurita-S\'anchez}, \citenamefont {Halevi},\ and\ \citenamefont
  {Cervantes-Gonz\'alez}}]{Zurita-Sanchez2009}  \BibitemOpen
  \bibfield  {author} {\bibinfo {author} {\bibfnamefont {J.~R.}\ \bibnamefont
  {Zurita-S\'anchez}}, \bibinfo {author} {\bibfnamefont {P.}~\bibnamefont
  {Halevi}},\ and\ \bibinfo {author} {\bibfnamefont {J.~C.}\ \bibnamefont
  {Cervantes-Gonz\'alez}},\ }\bibfield  {title} {\bibinfo {title} {Reflection
  and transmission of a wave incident on a slab with a time-periodic dielectric
  function $\epsilon(t)$},\ }\href {https://doi.org/10.1103/PhysRevA.79.053821}
  {\bibfield  {journal} {\bibinfo  {journal} {Phys. Rev. A}\ }\textbf {\bibinfo
  {volume} {79}},\ \bibinfo {pages} {053821} (\bibinfo {year}
  {2009})}\BibitemShut {NoStop}\bibitem [{\citenamefont {Reyes-Ayona}\ and\ \citenamefont
  {Halevi}(2015)}]{Reyes-Ayona2015}  \BibitemOpen
  \bibfield  {author} {\bibinfo {author} {\bibfnamefont {J.~R.}\ \bibnamefont
  {Reyes-Ayona}}\ and\ \bibinfo {author} {\bibfnamefont {P.}~\bibnamefont
  {Halevi}},\ }\bibfield  {title} {\bibinfo {title} {Observation of genuine
  wave vector ($k$ or $\beta$) gap in a dynamic transmission line and temporal
  photonic crystals},\ }\href {https://doi.org/10.1063/1.4928659} {\bibfield
  {journal} {\bibinfo  {journal} {Appl. Phys. Lett.}\ }\textbf {\bibinfo
  {volume} {107}},\ \bibinfo {pages} {074101} (\bibinfo {year}
  {2015})}\BibitemShut {NoStop}\bibitem [{\citenamefont {Lyubarov}\ \emph {et~al.}(2022)\citenamefont
  {Lyubarov}, \citenamefont {Lumer}, \citenamefont {Dikopoltsev}, \citenamefont
  {Lustig}, \citenamefont {Sharabi},\ and\ \citenamefont
  {Segev}}]{Lyubarov2022}  \BibitemOpen
  \bibfield  {author} {\bibinfo {author} {\bibfnamefont {M.}~\bibnamefont
  {Lyubarov}}, \bibinfo {author} {\bibfnamefont {Y.}~\bibnamefont {Lumer}},
  \bibinfo {author} {\bibfnamefont {A.}~\bibnamefont {Dikopoltsev}}, \bibinfo
  {author} {\bibfnamefont {E.}~\bibnamefont {Lustig}}, \bibinfo {author}
  {\bibfnamefont {Y.}~\bibnamefont {Sharabi}},\ and\ \bibinfo {author}
  {\bibfnamefont {M.}~\bibnamefont {Segev}},\ }\bibfield  {title} {\bibinfo
  {title} {{Amplified emission and lasing in photonic time crystals}},\ }\href
  {https://doi.org/10.1126/science.abo3324} {\bibfield  {journal} {\bibinfo
  {journal} {Science}\ }\textbf {\bibinfo {volume} {377}},\ \bibinfo {pages}
  {425} (\bibinfo {year} {2022})}\BibitemShut {NoStop}\bibitem [{\citenamefont {Dikopoltsev}\ \emph {et~al.}(2022)\citenamefont
  {Dikopoltsev}, \citenamefont {Sharabi}, \citenamefont {Lyubarov},
  \citenamefont {Lumer}, \citenamefont {Tsesses}, \citenamefont {Lustig},
  \citenamefont {Kaminer},\ and\ \citenamefont {Segev}}]{Dikopoltsev2022}  \BibitemOpen
  \bibfield  {author} {\bibinfo {author} {\bibfnamefont {A.}~\bibnamefont
  {Dikopoltsev}}, \bibinfo {author} {\bibfnamefont {Y.}~\bibnamefont
  {Sharabi}}, \bibinfo {author} {\bibfnamefont {M.}~\bibnamefont {Lyubarov}},
  \bibinfo {author} {\bibfnamefont {Y.}~\bibnamefont {Lumer}}, \bibinfo
  {author} {\bibfnamefont {S.}~\bibnamefont {Tsesses}}, \bibinfo {author}
  {\bibfnamefont {E.}~\bibnamefont {Lustig}}, \bibinfo {author} {\bibfnamefont
  {I.}~\bibnamefont {Kaminer}},\ and\ \bibinfo {author} {\bibfnamefont
  {M.}~\bibnamefont {Segev}},\ }\bibfield  {title} {\bibinfo {title} {{Light
  emission by free electrons in photonic time-crystals}},\ }\href
  {https://doi.org/10.1073/pnas.2119705119} {\bibfield  {journal} {\bibinfo
  {journal} {Proc. Natl. Acad. Sci. USA}\ }\textbf {\bibinfo {volume} {119}},\
  \bibinfo {pages} {e2119705119} (\bibinfo {year} {2022})}\BibitemShut
  {NoStop}\bibitem [{\citenamefont {Lee}\ \emph {et~al.}(2026)\citenamefont {Lee},
  \citenamefont {Kyung}, \citenamefont {Kim}, \citenamefont {Park},
  \citenamefont {Lee}, \citenamefont {Choi}, \citenamefont {Chan},
  \citenamefont {Shin}, \citenamefont {Kim},\ and\ \citenamefont
  {Min}}]{Lee2026PRL}  \BibitemOpen
  \bibfield  {author} {\bibinfo {author} {\bibfnamefont {K.}~\bibnamefont
  {Lee}}, \bibinfo {author} {\bibfnamefont {M.}~\bibnamefont {Kyung}}, \bibinfo
  {author} {\bibfnamefont {Y.}~\bibnamefont {Kim}}, \bibinfo {author}
  {\bibfnamefont {J.}~\bibnamefont {Park}}, \bibinfo {author} {\bibfnamefont
  {H.}~\bibnamefont {Lee}}, \bibinfo {author} {\bibfnamefont {J.}~\bibnamefont
  {Choi}}, \bibinfo {author} {\bibfnamefont {C.~T.}\ \bibnamefont {Chan}},
  \bibinfo {author} {\bibfnamefont {J.}~\bibnamefont {Shin}}, \bibinfo {author}
  {\bibfnamefont {K.~W.}\ \bibnamefont {Kim}},\ and\ \bibinfo {author}
  {\bibfnamefont {B.}~\bibnamefont {Min}},\ }\bibfield  {title} {\bibinfo
  {title} {{Analogs of Spontaneous Emission and Lasing in Photonic Time
  Crystals}},\ }\href {https://doi.org/10.1103/hh9h-qzpk} {\bibfield  {journal}
  {\bibinfo  {journal} {Phys. Rev. Lett.}\ }\textbf {\bibinfo {volume} {136}},\
  \bibinfo {pages} {093802} (\bibinfo {year} {2026})}\BibitemShut {NoStop}\bibitem [{\citenamefont {Carminati}\ \emph {et~al.}(2021)\citenamefont
  {Carminati}, \citenamefont {Chen}, \citenamefont {Pierrat},\ and\
  \citenamefont {Shapiro}}]{Carminati2021}  \BibitemOpen
  \bibfield  {author} {\bibinfo {author} {\bibfnamefont {R.}~\bibnamefont
  {Carminati}}, \bibinfo {author} {\bibfnamefont {H.}~\bibnamefont {Chen}},
  \bibinfo {author} {\bibfnamefont {R.}~\bibnamefont {Pierrat}},\ and\ \bibinfo
  {author} {\bibfnamefont {B.}~\bibnamefont {Shapiro}},\ }\bibfield  {title}
  {\bibinfo {title} {{Universal Statistics of Waves in a Random Time-Varying
  Medium}},\ }\href {https://doi.org/10.1103/physrevlett.127.094101} {\bibfield
   {journal} {\bibinfo  {journal} {Phys. Rev. Lett.}\ }\textbf {\bibinfo
  {volume} {127}},\ \bibinfo {pages} {094101} (\bibinfo {year}
  {2021})}\BibitemShut {NoStop}\bibitem [{\citenamefont {Apffel}\ \emph {et~al.}(2022)\citenamefont {Apffel},
  \citenamefont {Wildeman}, \citenamefont {Eddi},\ and\ \citenamefont
  {Fort}}]{Apffel2022}  \BibitemOpen
  \bibfield  {author} {\bibinfo {author} {\bibfnamefont {B.}~\bibnamefont
  {Apffel}}, \bibinfo {author} {\bibfnamefont {S.}~\bibnamefont {Wildeman}},
  \bibinfo {author} {\bibfnamefont {A.}~\bibnamefont {Eddi}},\ and\ \bibinfo
  {author} {\bibfnamefont {E.}~\bibnamefont {Fort}},\ }\bibfield  {title}
  {\bibinfo {title} {{Experimental Implementation of Wave Propagation in
  Disordered Time-Varying Media}},\ }\href
  {https://doi.org/10.1103/physrevlett.128.094503} {\bibfield  {journal}
  {\bibinfo  {journal} {Phys. Rev. Lett.}\ }\textbf {\bibinfo {volume} {128}},\
  \bibinfo {pages} {094503} (\bibinfo {year} {2022})}\BibitemShut {NoStop}\bibitem [{\citenamefont {Sharabi}\ \emph {et~al.}(2021)\citenamefont
  {Sharabi}, \citenamefont {Lustig},\ and\ \citenamefont
  {Segev}}]{Sharabi2021}  \BibitemOpen
  \bibfield  {author} {\bibinfo {author} {\bibfnamefont {Y.}~\bibnamefont
  {Sharabi}}, \bibinfo {author} {\bibfnamefont {E.}~\bibnamefont {Lustig}},\
  and\ \bibinfo {author} {\bibfnamefont {M.}~\bibnamefont {Segev}},\ }\bibfield
   {title} {\bibinfo {title} {{Disordered Photonic Time Crystals}},\ }\href
  {https://doi.org/10.1103/physrevlett.126.163902} {\bibfield  {journal}
  {\bibinfo  {journal} {Phys. Rev. Lett.}\ }\textbf {\bibinfo {volume} {126}},\
  \bibinfo {pages} {163902} (\bibinfo {year} {2021})}\BibitemShut {NoStop}\bibitem [{\citenamefont {Eswaran}\ \emph {et~al.}(2025)\citenamefont
  {Eswaran}, \citenamefont {Kopaei},\ and\ \citenamefont
  {Sacha}}]{Eswaran2025}  \BibitemOpen
  \bibfield  {author} {\bibinfo {author} {\bibfnamefont {K.~S.}\ \bibnamefont
  {Eswaran}}, \bibinfo {author} {\bibfnamefont {A.~E.}\ \bibnamefont
  {Kopaei}},\ and\ \bibinfo {author} {\bibfnamefont {K.}~\bibnamefont
  {Sacha}},\ }\bibfield  {title} {\bibinfo {title} {{Anderson localization in
  photonic time crystals}},\ }\href
  {https://doi.org/10.1103/physrevb.111.l180201} {\bibfield  {journal}
  {\bibinfo  {journal} {Phys. Rev. B}\ }\textbf {\bibinfo {volume} {111}},\
  \bibinfo {pages} {L180201} (\bibinfo {year} {2025})}\BibitemShut {NoStop}\bibitem [{\citenamefont {Sacha}\ and\ \citenamefont
  {Delande}(2016)}]{Sacha2016}  \BibitemOpen
  \bibfield  {author} {\bibinfo {author} {\bibfnamefont {K.}~\bibnamefont
  {Sacha}}\ and\ \bibinfo {author} {\bibfnamefont {D.}~\bibnamefont
  {Delande}},\ }\bibfield  {title} {\bibinfo {title} {{Anderson localization in
  the time domain}},\ }\href {https://doi.org/10.1103/physreva.94.023633}
  {\bibfield  {journal} {\bibinfo  {journal} {Phys. Rev. A}\ }\textbf {\bibinfo
  {volume} {94}},\ \bibinfo {pages} {023633} (\bibinfo {year}
  {2016})}\BibitemShut {NoStop}\bibitem [{\citenamefont {Delande}\ \emph {et~al.}(2017)\citenamefont
  {Delande}, \citenamefont {Morales-Molina},\ and\ \citenamefont
  {Sacha}}]{Delande2017}  \BibitemOpen
  \bibfield  {author} {\bibinfo {author} {\bibfnamefont {D.}~\bibnamefont
  {Delande}}, \bibinfo {author} {\bibfnamefont {L.}~\bibnamefont
  {Morales-Molina}},\ and\ \bibinfo {author} {\bibfnamefont {K.}~\bibnamefont
  {Sacha}},\ }\bibfield  {title} {\bibinfo {title} {{Three-Dimensional
  Localized-Delocalized Anderson Transition in the Time Domain}},\ }\href
  {https://doi.org/10.1103/physrevlett.119.230404} {\bibfield  {journal}
  {\bibinfo  {journal} {Phys. Rev. Lett.}\ }\textbf {\bibinfo {volume} {119}},\
  \bibinfo {pages} {230404} (\bibinfo {year} {2017})}\BibitemShut {NoStop}\bibitem [{\citenamefont {Anderson}(1958)}]{Anderson1958}  \BibitemOpen
  \bibfield  {author} {\bibinfo {author} {\bibfnamefont {P.~W.}\ \bibnamefont
  {Anderson}},\ }\bibfield  {title} {\bibinfo {title} {Absence of diffusion in
  certain random lattices},\ }\href {https://doi.org/10.1103/PhysRev.109.1492}
  {\bibfield  {journal} {\bibinfo  {journal} {Phys. Rev.}\ }\textbf {\bibinfo
  {volume} {109}},\ \bibinfo {pages} {1492} (\bibinfo {year}
  {1958})}\BibitemShut {NoStop}\bibitem [{\citenamefont {Lagendijk}\ \emph {et~al.}(2009)\citenamefont
  {Lagendijk}, \citenamefont {van Tiggelen},\ and\ \citenamefont
  {Wiersma}}]{Lagendijk2009}  \BibitemOpen
  \bibfield  {author} {\bibinfo {author} {\bibfnamefont {A.}~\bibnamefont
  {Lagendijk}}, \bibinfo {author} {\bibfnamefont {B.}~\bibnamefont {van
  Tiggelen}},\ and\ \bibinfo {author} {\bibfnamefont {D.~S.}\ \bibnamefont
  {Wiersma}},\ }\bibfield  {title} {\bibinfo {title} {Fifty years of {Anderson}
  localization},\ }\href {https://doi.org/10.1063/1.3206091} {\bibfield
  {journal} {\bibinfo  {journal} {Phys. Today}\ }\textbf {\bibinfo {volume}
  {62}},\ \bibinfo {pages} {24} (\bibinfo {year} {2009})}\BibitemShut {NoStop}\bibitem [{\citenamefont {Billy}\ \emph {et~al.}(2008)\citenamefont {Billy},
  \citenamefont {Josse}, \citenamefont {Zuo}, \citenamefont {Bernard},
  \citenamefont {Hambrecht}, \citenamefont {Lugan}, \citenamefont {Cl\'ement},
  \citenamefont {Sanchez-Palencia}, \citenamefont {Bouyer},\ and\ \citenamefont
  {Aspect}}]{Billy2008}  \BibitemOpen
  \bibfield  {author} {\bibinfo {author} {\bibfnamefont {J.}~\bibnamefont
  {Billy}}, \bibinfo {author} {\bibfnamefont {V.}~\bibnamefont {Josse}},
  \bibinfo {author} {\bibfnamefont {Z.}~\bibnamefont {Zuo}}, \bibinfo {author}
  {\bibfnamefont {A.}~\bibnamefont {Bernard}}, \bibinfo {author} {\bibfnamefont
  {B.}~\bibnamefont {Hambrecht}}, \bibinfo {author} {\bibfnamefont
  {P.}~\bibnamefont {Lugan}}, \bibinfo {author} {\bibfnamefont
  {D.}~\bibnamefont {Cl\'ement}}, \bibinfo {author} {\bibfnamefont
  {L.}~\bibnamefont {Sanchez-Palencia}}, \bibinfo {author} {\bibfnamefont
  {P.}~\bibnamefont {Bouyer}},\ and\ \bibinfo {author} {\bibfnamefont
  {A.}~\bibnamefont {Aspect}},\ }\bibfield  {title} {\bibinfo {title} {Direct
  observation of {Anderson} localization of matter waves in a controlled
  disorder},\ }\href {https://doi.org/10.1038/nature07000} {\bibfield
  {journal} {\bibinfo  {journal} {Nature}\ }\textbf {\bibinfo {volume} {453}},\
  \bibinfo {pages} {891} (\bibinfo {year} {2008})}\BibitemShut {NoStop}\bibitem [{\citenamefont {Lahini}\ \emph {et~al.}(2008)\citenamefont {Lahini},
  \citenamefont {Avidan}, \citenamefont {Pozzi}, \citenamefont {Sorel},
  \citenamefont {Morandotti}, \citenamefont {Christodoulides},\ and\
  \citenamefont {Silberberg}}]{Lahini2008}  \BibitemOpen
  \bibfield  {author} {\bibinfo {author} {\bibfnamefont {Y.}~\bibnamefont
  {Lahini}}, \bibinfo {author} {\bibfnamefont {A.}~\bibnamefont {Avidan}},
  \bibinfo {author} {\bibfnamefont {F.}~\bibnamefont {Pozzi}}, \bibinfo
  {author} {\bibfnamefont {M.}~\bibnamefont {Sorel}}, \bibinfo {author}
  {\bibfnamefont {R.}~\bibnamefont {Morandotti}}, \bibinfo {author}
  {\bibfnamefont {D.~N.}\ \bibnamefont {Christodoulides}},\ and\ \bibinfo
  {author} {\bibfnamefont {Y.}~\bibnamefont {Silberberg}},\ }\bibfield  {title}
  {\bibinfo {title} {Anderson localization and nonlinearity in one-dimensional
  disordered photonic lattices},\ }\href
  {https://doi.org/10.1103/PhysRevLett.100.013906} {\bibfield  {journal}
  {\bibinfo  {journal} {Phys. Rev. Lett.}\ }\textbf {\bibinfo {volume} {100}},\
  \bibinfo {pages} {013906} (\bibinfo {year} {2008})}\BibitemShut {NoStop}\bibitem [{\citenamefont {Furstenberg}(1963)}]{Furstenberg1963}  \BibitemOpen
  \bibfield  {author} {\bibinfo {author} {\bibfnamefont {H.}~\bibnamefont
  {Furstenberg}},\ }\bibfield  {title} {\bibinfo {title} {Noncommuting random
  products},\ }\href {https://doi.org/10.1090/S0002-9947-1963-0163345-0}
  {\bibfield  {journal} {\bibinfo  {journal} {Trans. Am. Math. Soc.}\ }\textbf
  {\bibinfo {volume} {108}},\ \bibinfo {pages} {377} (\bibinfo {year}
  {1963})}\BibitemShut {NoStop}\bibitem [{\citenamefont {Tessieri}\ and\ \citenamefont
  {Izrailev}(2000)}]{TessieriIzrailev1999}  \BibitemOpen
  \bibfield  {author} {\bibinfo {author} {\bibfnamefont {L.}~\bibnamefont
  {Tessieri}}\ and\ \bibinfo {author} {\bibfnamefont {F.~M.}\ \bibnamefont
  {Izrailev}},\ }\bibfield  {title} {\bibinfo {title} {Anderson localization as
  a parametric instability of the linear kicked oscillator},\ }\href
  {https://doi.org/10.1103/PhysRevE.62.3090} {\bibfield  {journal} {\bibinfo
  {journal} {Phys. Rev. E}\ }\textbf {\bibinfo {volume} {62}},\ \bibinfo
  {pages} {3090} (\bibinfo {year} {2000})}\BibitemShut {NoStop}\bibitem [{\citenamefont {Mallick}\ and\ \citenamefont
  {Marcq}(2002)}]{MallickMarcq2002}  \BibitemOpen
  \bibfield  {author} {\bibinfo {author} {\bibfnamefont {K.}~\bibnamefont
  {Mallick}}\ and\ \bibinfo {author} {\bibfnamefont {P.}~\bibnamefont
  {Marcq}},\ }\bibfield  {title} {\bibinfo {title} {Anomalous diffusion in
  nonlinear oscillators with multiplicative noise},\ }\href
  {https://doi.org/10.1103/PhysRevE.66.041113} {\bibfield  {journal} {\bibinfo
  {journal} {Phys. Rev. E}\ }\textbf {\bibinfo {volume} {66}},\ \bibinfo
  {pages} {041113} (\bibinfo {year} {2002})}\BibitemShut {NoStop}\bibitem [{\citenamefont {Aubry}\ and\ \citenamefont
  {Andr{\'e}}(1980)}]{AubryAndre1980}  \BibitemOpen
  \bibfield  {author} {\bibinfo {author} {\bibfnamefont {S.}~\bibnamefont
  {Aubry}}\ and\ \bibinfo {author} {\bibfnamefont {G.}~\bibnamefont
  {Andr{\'e}}},\ }\bibfield  {title} {\bibinfo {title} {{Analyticity breaking
  and Anderson localization in incommensurate lattices}},\ }in\ \href@noop {}
  {\emph {\bibinfo {booktitle} {Group Theoretical Methods in Physics}}},\
  \bibinfo {series} {Annals of the Israel Physical Society}, Vol.~\bibinfo
  {volume} {3},\ \bibinfo {editor} {edited by\ \bibinfo {editor} {\bibfnamefont
  {L.~P.}\ \bibnamefont {Horowitz}}\ and\ \bibinfo {editor} {\bibfnamefont
  {Y.}~\bibnamefont {Ne'eman}}}\ (\bibinfo  {publisher} {Adam Hilger and The
  Israel Physical Society},\ \bibinfo {address} {Bristol},\ \bibinfo {year}
  {1980})\ pp.\ \bibinfo {pages} {133--164}\BibitemShut {NoStop}\bibitem [{\citenamefont {Jitomirskaya}(1999)}]{Jitomirskaya1999}  \BibitemOpen
  \bibfield  {author} {\bibinfo {author} {\bibfnamefont {S.~Y.}\ \bibnamefont
  {Jitomirskaya}},\ }\bibfield  {title} {\bibinfo {title} {{Metal-insulator
  transition for the almost Mathieu operator}},\ }\href
  {https://doi.org/10.2307/121066} {\bibfield  {journal} {\bibinfo  {journal}
  {Ann. Math.}\ }\textbf {\bibinfo {volume} {150}},\ \bibinfo {pages} {1159}
  (\bibinfo {year} {1999})}\BibitemShut {NoStop}\bibitem [{\citenamefont {Roati}\ \emph {et~al.}(2008)\citenamefont {Roati},
  \citenamefont {D'Errico}, \citenamefont {Fallani}, \citenamefont {Fattori},
  \citenamefont {Fort}, \citenamefont {Zaccanti}, \citenamefont {Modugno},
  \citenamefont {Modugno},\ and\ \citenamefont {Inguscio}}]{Roati2008}  \BibitemOpen
  \bibfield  {author} {\bibinfo {author} {\bibfnamefont {G.}~\bibnamefont
  {Roati}}, \bibinfo {author} {\bibfnamefont {C.}~\bibnamefont {D'Errico}},
  \bibinfo {author} {\bibfnamefont {L.}~\bibnamefont {Fallani}}, \bibinfo
  {author} {\bibfnamefont {M.}~\bibnamefont {Fattori}}, \bibinfo {author}
  {\bibfnamefont {C.}~\bibnamefont {Fort}}, \bibinfo {author} {\bibfnamefont
  {M.}~\bibnamefont {Zaccanti}}, \bibinfo {author} {\bibfnamefont
  {G.}~\bibnamefont {Modugno}}, \bibinfo {author} {\bibfnamefont
  {M.}~\bibnamefont {Modugno}},\ and\ \bibinfo {author} {\bibfnamefont
  {M.}~\bibnamefont {Inguscio}},\ }\bibfield  {title} {\bibinfo {title}
  {{Anderson localization of a non-interacting Bose--Einstein condensate}},\
  }\href {https://doi.org/10.1038/nature07071} {\bibfield  {journal} {\bibinfo
  {journal} {Nature}\ }\textbf {\bibinfo {volume} {453}},\ \bibinfo {pages}
  {895} (\bibinfo {year} {2008})}\BibitemShut {NoStop}\bibitem [{\citenamefont {Lahini}\ \emph {et~al.}(2009)\citenamefont {Lahini},
  \citenamefont {Pugatch}, \citenamefont {Pozzi}, \citenamefont {Sorel},
  \citenamefont {Morandotti}, \citenamefont {Davidson},\ and\ \citenamefont
  {Silberberg}}]{Lahini2009}  \BibitemOpen
  \bibfield  {author} {\bibinfo {author} {\bibfnamefont {Y.}~\bibnamefont
  {Lahini}}, \bibinfo {author} {\bibfnamefont {R.}~\bibnamefont {Pugatch}},
  \bibinfo {author} {\bibfnamefont {F.}~\bibnamefont {Pozzi}}, \bibinfo
  {author} {\bibfnamefont {M.}~\bibnamefont {Sorel}}, \bibinfo {author}
  {\bibfnamefont {R.}~\bibnamefont {Morandotti}}, \bibinfo {author}
  {\bibfnamefont {N.}~\bibnamefont {Davidson}},\ and\ \bibinfo {author}
  {\bibfnamefont {Y.}~\bibnamefont {Silberberg}},\ }\bibfield  {title}
  {\bibinfo {title} {{Observation of a Localization Transition in Quasiperiodic
  Photonic Lattices}},\ }\href {https://doi.org/10.1103/physrevlett.103.013901}
  {\bibfield  {journal} {\bibinfo  {journal} {Phys. Rev. Lett.}\ }\textbf
  {\bibinfo {volume} {103}},\ \bibinfo {pages} {013901} (\bibinfo {year}
  {2009})}\BibitemShut {NoStop}\bibitem [{\citenamefont {Biddle}\ and\ \citenamefont
  {Das~Sarma}(2010)}]{BiddleDasSarma2010}  \BibitemOpen
  \bibfield  {author} {\bibinfo {author} {\bibfnamefont {J.}~\bibnamefont
  {Biddle}}\ and\ \bibinfo {author} {\bibfnamefont {S.}~\bibnamefont
  {Das~Sarma}},\ }\bibfield  {title} {\bibinfo {title} {{Predicted Mobility
  Edges in One-Dimensional Incommensurate Optical Lattices: An Exactly Solvable
  Model of Anderson Localization}},\ }\href
  {https://doi.org/10.1103/physrevlett.104.070601} {\bibfield  {journal}
  {\bibinfo  {journal} {Phys. Rev. Lett.}\ }\textbf {\bibinfo {volume} {104}},\
  \bibinfo {pages} {070601} (\bibinfo {year} {2010})}\BibitemShut {NoStop}\bibitem [{\citenamefont {Ganeshan}\ \emph {et~al.}(2015)\citenamefont
  {Ganeshan}, \citenamefont {Pixley},\ and\ \citenamefont
  {Das~Sarma}}]{Ganeshan2015}  \BibitemOpen
  \bibfield  {author} {\bibinfo {author} {\bibfnamefont {S.}~\bibnamefont
  {Ganeshan}}, \bibinfo {author} {\bibfnamefont {J.~H.}\ \bibnamefont
  {Pixley}},\ and\ \bibinfo {author} {\bibfnamefont {S.}~\bibnamefont
  {Das~Sarma}},\ }\bibfield  {title} {\bibinfo {title} {{Nearest Neighbor Tight
  Binding Models with an Exact Mobility Edge in One Dimension}},\ }\href
  {https://doi.org/10.1103/physrevlett.114.146601} {\bibfield  {journal}
  {\bibinfo  {journal} {Phys. Rev. Lett.}\ }\textbf {\bibinfo {volume} {114}},\
  \bibinfo {pages} {146601} (\bibinfo {year} {2015})}\BibitemShut {NoStop}\bibitem [{\citenamefont {L{\"u}schen}\ \emph {et~al.}(2018)\citenamefont
  {L{\"u}schen}, \citenamefont {Scherg}, \citenamefont {Kohlert}, \citenamefont
  {Schreiber}, \citenamefont {Bordia}, \citenamefont {Li}, \citenamefont
  {Das~Sarma},\ and\ \citenamefont {Bloch}}]{Luschen2018}  \BibitemOpen
  \bibfield  {author} {\bibinfo {author} {\bibfnamefont {H.~P.}\ \bibnamefont
  {L{\"u}schen}}, \bibinfo {author} {\bibfnamefont {S.}~\bibnamefont {Scherg}},
  \bibinfo {author} {\bibfnamefont {T.}~\bibnamefont {Kohlert}}, \bibinfo
  {author} {\bibfnamefont {M.}~\bibnamefont {Schreiber}}, \bibinfo {author}
  {\bibfnamefont {P.}~\bibnamefont {Bordia}}, \bibinfo {author} {\bibfnamefont
  {X.}~\bibnamefont {Li}}, \bibinfo {author} {\bibfnamefont {S.}~\bibnamefont
  {Das~Sarma}},\ and\ \bibinfo {author} {\bibfnamefont {I.}~\bibnamefont
  {Bloch}},\ }\bibfield  {title} {\bibinfo {title} {{Single-Particle Mobility
  Edge in a One-Dimensional Quasiperiodic Optical Lattice}},\ }\href
  {https://doi.org/10.1103/physrevlett.120.160404} {\bibfield  {journal}
  {\bibinfo  {journal} {Phys. Rev. Lett.}\ }\textbf {\bibinfo {volume} {120}},\
  \bibinfo {pages} {160404} (\bibinfo {year} {2018})}\BibitemShut {NoStop}\bibitem [{\citenamefont {An}\ \emph {et~al.}(2021)\citenamefont {An},
  \citenamefont {Padavi{\'c}}, \citenamefont {Meier}, \citenamefont {Hegde},
  \citenamefont {Ganeshan}, \citenamefont {Pixley}, \citenamefont
  {Vishveshwara},\ and\ \citenamefont {Gadway}}]{An2021}  \BibitemOpen
  \bibfield  {author} {\bibinfo {author} {\bibfnamefont {F.~A.}\ \bibnamefont
  {An}}, \bibinfo {author} {\bibfnamefont {K.}~\bibnamefont {Padavi{\'c}}},
  \bibinfo {author} {\bibfnamefont {E.~J.}\ \bibnamefont {Meier}}, \bibinfo
  {author} {\bibfnamefont {S.}~\bibnamefont {Hegde}}, \bibinfo {author}
  {\bibfnamefont {S.}~\bibnamefont {Ganeshan}}, \bibinfo {author}
  {\bibfnamefont {J.~H.}\ \bibnamefont {Pixley}}, \bibinfo {author}
  {\bibfnamefont {S.}~\bibnamefont {Vishveshwara}},\ and\ \bibinfo {author}
  {\bibfnamefont {B.}~\bibnamefont {Gadway}},\ }\bibfield  {title} {\bibinfo
  {title} {{Interactions and Mobility Edges: Observing the Generalized
  Aubry-Andr{\'e} Model}},\ }\href
  {https://doi.org/10.1103/physrevlett.126.040603} {\bibfield  {journal}
  {\bibinfo  {journal} {Phys. Rev. Lett.}\ }\textbf {\bibinfo {volume} {126}},\
  \bibinfo {pages} {040603} (\bibinfo {year} {2021})}\BibitemShut {NoStop}\bibitem [{\citenamefont {Avila}(2015)}]{Avila2015}  \BibitemOpen
  \bibfield  {author} {\bibinfo {author} {\bibfnamefont {A.}~\bibnamefont
  {Avila}},\ }\bibfield  {title} {\bibinfo {title} {{Global theory of
  one-frequency Schr{\"o}dinger operators}},\ }\href
  {https://doi.org/10.1007/s11511-015-0128-7} {\bibfield  {journal} {\bibinfo
  {journal} {Acta Math.}\ }\textbf {\bibinfo {volume} {215}},\ \bibinfo {pages}
  {1} (\bibinfo {year} {2015})}\BibitemShut {NoStop}\bibitem [{\citenamefont {Wang}\ \emph {et~al.}(2020)\citenamefont {Wang},
  \citenamefont {Xia}, \citenamefont {Zhang}, \citenamefont {Yao},
  \citenamefont {Chen}, \citenamefont {You}, \citenamefont {Zhou},\ and\
  \citenamefont {Liu}}]{Wang2020mosaic}  \BibitemOpen
  \bibfield  {author} {\bibinfo {author} {\bibfnamefont {Y.}~\bibnamefont
  {Wang}}, \bibinfo {author} {\bibfnamefont {X.}~\bibnamefont {Xia}}, \bibinfo
  {author} {\bibfnamefont {L.}~\bibnamefont {Zhang}}, \bibinfo {author}
  {\bibfnamefont {H.}~\bibnamefont {Yao}}, \bibinfo {author} {\bibfnamefont
  {S.}~\bibnamefont {Chen}}, \bibinfo {author} {\bibfnamefont {J.}~\bibnamefont
  {You}}, \bibinfo {author} {\bibfnamefont {Q.}~\bibnamefont {Zhou}},\ and\
  \bibinfo {author} {\bibfnamefont {X.-J.}\ \bibnamefont {Liu}},\ }\bibfield
  {title} {\bibinfo {title} {{One-Dimensional Quasiperiodic Mosaic Lattice with
  Exact Mobility Edges}},\ }\href
  {https://doi.org/10.1103/physrevlett.125.196604} {\bibfield  {journal}
  {\bibinfo  {journal} {Phys. Rev. Lett.}\ }\textbf {\bibinfo {volume} {125}},\
  \bibinfo {pages} {196604} (\bibinfo {year} {2020})}\BibitemShut {NoStop}\bibitem [{\citenamefont {Wang}\ \emph {et~al.}(2023)\citenamefont {Wang},
  \citenamefont {Xia}, \citenamefont {You}, \citenamefont {Zheng},\ and\
  \citenamefont {Zhou}}]{WangXiaYouZhengZhou2023}  \BibitemOpen
  \bibfield  {author} {\bibinfo {author} {\bibfnamefont {Y.}~\bibnamefont
  {Wang}}, \bibinfo {author} {\bibfnamefont {X.}~\bibnamefont {Xia}}, \bibinfo
  {author} {\bibfnamefont {J.}~\bibnamefont {You}}, \bibinfo {author}
  {\bibfnamefont {Z.}~\bibnamefont {Zheng}},\ and\ \bibinfo {author}
  {\bibfnamefont {Q.}~\bibnamefont {Zhou}},\ }\bibfield  {title} {\bibinfo
  {title} {{Exact Mobility Edges for 1D Quasiperiodic Models}},\ }\href
  {https://doi.org/10.1007/s00220-023-04695-9} {\bibfield  {journal} {\bibinfo
  {journal} {Commun. Math. Phys.}\ }\textbf {\bibinfo {volume} {401}},\
  \bibinfo {pages} {2521} (\bibinfo {year} {2023})}\BibitemShut {NoStop}\bibitem [{\citenamefont {Casati}\ \emph {et~al.}(1989)\citenamefont {Casati},
  \citenamefont {Guarneri},\ and\ \citenamefont {Shepelyansky}}]{Casati1989}  \BibitemOpen
  \bibfield  {author} {\bibinfo {author} {\bibfnamefont {G.}~\bibnamefont
  {Casati}}, \bibinfo {author} {\bibfnamefont {I.}~\bibnamefont {Guarneri}},\
  and\ \bibinfo {author} {\bibfnamefont {D.~L.}\ \bibnamefont {Shepelyansky}},\
  }\bibfield  {title} {\bibinfo {title} {{Anderson Transition in a
  One-Dimensional System with Three Incommensurate Frequencies}},\ }\href
  {https://doi.org/10.1103/PhysRevLett.62.345} {\bibfield  {journal} {\bibinfo
  {journal} {Phys. Rev. Lett.}\ }\textbf {\bibinfo {volume} {62}},\ \bibinfo
  {pages} {345} (\bibinfo {year} {1989})}\BibitemShut {NoStop}\bibitem [{\citenamefont {Chab{\'e}}\ \emph {et~al.}(2008)\citenamefont
  {Chab{\'e}}, \citenamefont {Lemari{\'e}}, \citenamefont {Gr{\'e}maud},
  \citenamefont {Delande}, \citenamefont {Szriftgiser},\ and\ \citenamefont
  {Garreau}}]{Chabe2008}  \BibitemOpen
  \bibfield  {author} {\bibinfo {author} {\bibfnamefont {J.}~\bibnamefont
  {Chab{\'e}}}, \bibinfo {author} {\bibfnamefont {G.}~\bibnamefont
  {Lemari{\'e}}}, \bibinfo {author} {\bibfnamefont {B.}~\bibnamefont
  {Gr{\'e}maud}}, \bibinfo {author} {\bibfnamefont {D.}~\bibnamefont
  {Delande}}, \bibinfo {author} {\bibfnamefont {P.}~\bibnamefont
  {Szriftgiser}},\ and\ \bibinfo {author} {\bibfnamefont {J.~C.}\ \bibnamefont
  {Garreau}},\ }\bibfield  {title} {\bibinfo {title} {{Experimental Observation
  of the Anderson Metal-Insulator Transition with Atomic Matter Waves}},\
  }\href {https://doi.org/10.1103/physrevlett.101.255702} {\bibfield  {journal}
  {\bibinfo  {journal} {Phys. Rev. Lett.}\ }\textbf {\bibinfo {volume} {101}},\
  \bibinfo {pages} {255702} (\bibinfo {year} {2008})}\BibitemShut {NoStop}\bibitem [{\citenamefont {Ruchiev}\ and\ \citenamefont
  {Davies}(2026)}]{Ruchiev2026}  \BibitemOpen
  \bibfield  {author} {\bibinfo {author} {\bibfnamefont {I.}~\bibnamefont
  {Ruchiev}}\ and\ \bibinfo {author} {\bibfnamefont {B.}~\bibnamefont
  {Davies}},\ }\bibfield  {title} {\bibinfo {title} {{Energy estimates for wave
  amplification in quasiperiodic Fibonacci time-modulated media}},\ }\href
  {https://doi.org/10.1103/w9bj-gmml} {\bibfield  {journal} {\bibinfo
  {journal} {Phys. Rev. A}\ }\textbf {\bibinfo {volume} {114}},\ \bibinfo
  {pages} {023502} (\bibinfo {year} {2026})}\BibitemShut {NoStop}\bibitem [{\citenamefont {Liu}\ \emph {et~al.}(2025)\citenamefont {Liu},
  \citenamefont {Chen}, \citenamefont {Cai}, \citenamefont {Ren},\ and\
  \citenamefont {Xu}}]{Liu2025OL}  \BibitemOpen
  \bibfield  {author} {\bibinfo {author} {\bibfnamefont {Q.}~\bibnamefont
  {Liu}}, \bibinfo {author} {\bibfnamefont {Z.}~\bibnamefont {Chen}}, \bibinfo
  {author} {\bibfnamefont {W.}~\bibnamefont {Cai}}, \bibinfo {author}
  {\bibfnamefont {M.}~\bibnamefont {Ren}},\ and\ \bibinfo {author}
  {\bibfnamefont {J.}~\bibnamefont {Xu}},\ }\bibfield  {title} {\bibinfo
  {title} {{Amplified transmission in full-k space empowered by photonic
  temporal quasicrystal}},\ }\href {https://doi.org/10.1364/OL.568304}
  {\bibfield  {journal} {\bibinfo  {journal} {Opt. Lett.}\ }\textbf {\bibinfo
  {volume} {50}},\ \bibinfo {pages} {5286} (\bibinfo {year}
  {2025})}\BibitemShut {NoStop}\bibitem [{\citenamefont {Coppolaro}\ \emph {et~al.}(2025)\citenamefont
  {Coppolaro}, \citenamefont {Moccia}, \citenamefont {Castaldi},\ and\
  \citenamefont {Galdi}}]{Coppolaro2025}  \BibitemOpen
  \bibfield  {author} {\bibinfo {author} {\bibfnamefont {M.}~\bibnamefont
  {Coppolaro}}, \bibinfo {author} {\bibfnamefont {M.}~\bibnamefont {Moccia}},
  \bibinfo {author} {\bibfnamefont {G.}~\bibnamefont {Castaldi}},\ and\
  \bibinfo {author} {\bibfnamefont {V.}~\bibnamefont {Galdi}},\ }\bibfield
  {title} {\bibinfo {title} {{Exploring aperiodic order in photonic time
  crystals}},\ }\href {https://doi.org/10.1103/vw6g-kq3z} {\bibfield  {journal}
  {\bibinfo  {journal} {Phys. Rev. Applied}\ }\textbf {\bibinfo {volume}
  {24}},\ \bibinfo {pages} {L011001} (\bibinfo {year} {2025})}\BibitemShut
  {NoStop}\bibitem [{\citenamefont {Ara{\'u}jo}\ \emph {et~al.}(2025)\citenamefont
  {Ara{\'u}jo}, \citenamefont {Costa}, \citenamefont {Azevedo}, \citenamefont
  {Viswanathan},\ and\ \citenamefont {Bezerra}}]{Araujo2025}  \BibitemOpen
  \bibfield  {author} {\bibinfo {author} {\bibfnamefont {A.~C.}\ \bibnamefont
  {Ara{\'u}jo}}, \bibinfo {author} {\bibfnamefont {C.~H.}\ \bibnamefont
  {Costa}}, \bibinfo {author} {\bibfnamefont {S.}~\bibnamefont {Azevedo}},
  \bibinfo {author} {\bibfnamefont {G.~M.}\ \bibnamefont {Viswanathan}},\ and\
  \bibinfo {author} {\bibfnamefont {C.~G.}\ \bibnamefont {Bezerra}},\
  }\bibfield  {title} {\bibinfo {title} {{Optical properties of 1D photonic
  time quasicrystals}},\ }\href {https://doi.org/10.1364/JOSAB.542001}
  {\bibfield  {journal} {\bibinfo  {journal} {J. Opt. Soc. Am. B}\ }\textbf
  {\bibinfo {volume} {42}},\ \bibinfo {pages} {949} (\bibinfo {year}
  {2025})}\BibitemShut {NoStop}\bibitem [{\citenamefont {Ni}\ \emph {et~al.}(2025)\citenamefont {Ni},
  \citenamefont {Yin}, \citenamefont {Li},\ and\ \citenamefont
  {Al{\`u}}}]{Ni2025}  \BibitemOpen
  \bibfield  {author} {\bibinfo {author} {\bibfnamefont {X.}~\bibnamefont
  {Ni}}, \bibinfo {author} {\bibfnamefont {S.}~\bibnamefont {Yin}}, \bibinfo
  {author} {\bibfnamefont {H.}~\bibnamefont {Li}},\ and\ \bibinfo {author}
  {\bibfnamefont {A.}~\bibnamefont {Al{\`u}}},\ }\bibfield  {title} {\bibinfo
  {title} {{Topological wave phenomena in photonic time quasicrystals}},\
  }\href {https://doi.org/10.1103/physrevb.111.125421} {\bibfield  {journal}
  {\bibinfo  {journal} {Phys. Rev. B}\ }\textbf {\bibinfo {volume} {111}},\
  \bibinfo {pages} {125421} (\bibinfo {year} {2025})}\BibitemShut {NoStop}\bibitem [{\citenamefont {Nemirovsky-Levy}\ \emph {et~al.}(2023)\citenamefont
  {Nemirovsky-Levy}, \citenamefont {Konforty}, \citenamefont {Segal},
  \citenamefont {Cohen}, \citenamefont {Lubarov}, \citenamefont {Plotnik},\
  and\ \citenamefont {Segev}}]{NemirovskyLevy2023}  \BibitemOpen
  \bibfield  {author} {\bibinfo {author} {\bibfnamefont {L.}~\bibnamefont
  {Nemirovsky-Levy}}, \bibinfo {author} {\bibfnamefont {N.}~\bibnamefont
  {Konforty}}, \bibinfo {author} {\bibfnamefont {O.}~\bibnamefont {Segal}},
  \bibinfo {author} {\bibfnamefont {M.-I.}\ \bibnamefont {Cohen}}, \bibinfo
  {author} {\bibfnamefont {M.}~\bibnamefont {Lubarov}}, \bibinfo {author}
  {\bibfnamefont {Y.}~\bibnamefont {Plotnik}},\ and\ \bibinfo {author}
  {\bibfnamefont {M.}~\bibnamefont {Segev}},\ }\bibfield  {title} {\bibinfo
  {title} {{Photonic Time Quasi-Crystals}},\ }in\ \href
  {https://doi.org/10.1364/CLEO_FS.2023.FTu4D.4} {\emph {\bibinfo {booktitle}
  {CLEO 2023}}}\ (\bibinfo  {publisher} {Optica Publishing Group},\ \bibinfo
  {year} {2023})\ p.\ \bibinfo {pages} {FTu4D.4}\BibitemShut {NoStop}\bibitem [{\citenamefont {Longhi}(2019)}]{Longhi2019}  \BibitemOpen
  \bibfield  {author} {\bibinfo {author} {\bibfnamefont {S.}~\bibnamefont
  {Longhi}},\ }\bibfield  {title} {\bibinfo {title} {{Topological Phase
  Transition in non-Hermitian Quasicrystals}},\ }\href
  {https://doi.org/10.1103/physrevlett.122.237601} {\bibfield  {journal}
  {\bibinfo  {journal} {Phys. Rev. Lett.}\ }\textbf {\bibinfo {volume} {122}},\
  \bibinfo {pages} {237601} (\bibinfo {year} {2019})}\BibitemShut {NoStop}\bibitem [{\citenamefont {Liu}\ \emph {et~al.}(2021)\citenamefont {Liu},
  \citenamefont {Zhou},\ and\ \citenamefont {Chen}}]{LiuZhouChen2021}  \BibitemOpen
  \bibfield  {author} {\bibinfo {author} {\bibfnamefont {Y.}~\bibnamefont
  {Liu}}, \bibinfo {author} {\bibfnamefont {Q.}~\bibnamefont {Zhou}},\ and\
  \bibinfo {author} {\bibfnamefont {S.}~\bibnamefont {Chen}},\ }\bibfield
  {title} {\bibinfo {title} {{Localization transition, spectrum structure, and
  winding numbers for one-dimensional non-Hermitian quasicrystals}},\ }\href
  {https://doi.org/10.1103/physrevb.104.024201} {\bibfield  {journal} {\bibinfo
   {journal} {Phys. Rev. B}\ }\textbf {\bibinfo {volume} {104}},\ \bibinfo
  {pages} {024201} (\bibinfo {year} {2021})}\BibitemShut {NoStop}\bibitem [{\citenamefont {Chen}\ \emph {et~al.}(2026)\citenamefont {Chen},
  \citenamefont {Zheng},\ and\ \citenamefont {Zhu}}]{Chen2026NHPTQC}  \BibitemOpen
  \bibfield  {author} {\bibinfo {author} {\bibfnamefont {S.}~\bibnamefont
  {Chen}}, \bibinfo {author} {\bibfnamefont {Z.}~\bibnamefont {Zheng}},\ and\
  \bibinfo {author} {\bibfnamefont {Q.}~\bibnamefont {Zhu}},\ }\href
  {https://arxiv.org/abs/2609.24327} {\bibinfo {title} {{Non-Hermitian photonic
  time quasicrystal}}} (\bibinfo {year} {2026}),\ \Eprint
  {https://arxiv.org/abs/2609.24327} {arXiv:2609.24327} \BibitemShut {NoStop}\bibitem [{\citenamefont {Bloch}(1929)}]{Bloch1929}  \BibitemOpen
  \bibfield  {author} {\bibinfo {author} {\bibfnamefont {F.}~\bibnamefont
  {Bloch}},\ }\bibfield  {title} {\bibinfo {title} {{\"U}ber die
  {Q}uantenmechanik der {E}lektronen in {K}ristallgittern},\ }\href
  {https://doi.org/10.1007/BF01339455} {\bibfield  {journal} {\bibinfo
  {journal} {Z. Phys.}\ }\textbf {\bibinfo {volume} {52}},\ \bibinfo {pages}
  {555} (\bibinfo {year} {1929})}\BibitemShut {NoStop}\bibitem [{\citenamefont {Bourgain}\ and\ \citenamefont
  {Jitomirskaya}(2002)}]{BourgainJitomirskaya2002}  \BibitemOpen
  \bibfield  {author} {\bibinfo {author} {\bibfnamefont {J.}~\bibnamefont
  {Bourgain}}\ and\ \bibinfo {author} {\bibfnamefont {S.}~\bibnamefont
  {Jitomirskaya}},\ }\bibfield  {title} {\bibinfo {title} {Continuity of the
  {L}yapunov exponent for quasiperiodic operators with analytic potential},\
  }\href {https://doi.org/10.1023/A:1019751801035} {\bibfield  {journal}
  {\bibinfo  {journal} {Journal of Statistical Physics}\ }\textbf {\bibinfo
  {volume} {108}},\ \bibinfo {pages} {1203} (\bibinfo {year}
  {2002})}\BibitemShut {NoStop}\bibitem [{\citenamefont {Avila}\ and\ \citenamefont
  {Jitomirskaya}(2010)}]{AvilaJitomirskaya2010}  \BibitemOpen
  \bibfield  {author} {\bibinfo {author} {\bibfnamefont {A.}~\bibnamefont
  {Avila}}\ and\ \bibinfo {author} {\bibfnamefont {S.}~\bibnamefont
  {Jitomirskaya}},\ }\bibfield  {title} {\bibinfo {title} {{Almost localization
  and almost reducibility}},\ }\href {https://doi.org/10.4171/jems/191}
  {\bibfield  {journal} {\bibinfo  {journal} {J. Eur. Math. Soc.}\ }\textbf
  {\bibinfo {volume} {12}},\ \bibinfo {pages} {93} (\bibinfo {year}
  {2010})}\BibitemShut {NoStop}\bibitem [{\citenamefont {Broer}\ \emph {et~al.}(2003)\citenamefont {Broer},
  \citenamefont {Puig},\ and\ \citenamefont {Sim{\'o}}}]{BroerPuigSimo2003}  \BibitemOpen
  \bibfield  {author} {\bibinfo {author} {\bibfnamefont {H.}~\bibnamefont
  {Broer}}, \bibinfo {author} {\bibfnamefont {J.}~\bibnamefont {Puig}},\ and\
  \bibinfo {author} {\bibfnamefont {C.}~\bibnamefont {Sim{\'o}}},\ }\bibfield
  {title} {\bibinfo {title} {{Resonance Tongues and Instability Pockets in the
  Quasi--Periodic Hill--Schr{\"o}dinger Equation}},\ }\href
  {https://doi.org/10.1007/s00220-003-0935-0} {\bibfield  {journal} {\bibinfo
  {journal} {Commun. Math. Phys.}\ }\textbf {\bibinfo {volume} {241}},\
  \bibinfo {pages} {467} (\bibinfo {year} {2003})}\BibitemShut {NoStop}\bibitem [{\citenamefont {Jitomirskaya}\ \emph {et~al.}(2009)\citenamefont
  {Jitomirskaya}, \citenamefont {Koslover},\ and\ \citenamefont
  {Schulteis}}]{JKS2009}  \BibitemOpen
  \bibfield  {author} {\bibinfo {author} {\bibfnamefont {S.}~\bibnamefont
  {Jitomirskaya}}, \bibinfo {author} {\bibfnamefont {D.~A.}\ \bibnamefont
  {Koslover}},\ and\ \bibinfo {author} {\bibfnamefont {M.~S.}\ \bibnamefont
  {Schulteis}},\ }\bibfield  {title} {\bibinfo {title} {Continuity of the
  {L}yapunov exponent for analytic quasiperiodic cocycles},\ }\href
  {https://doi.org/10.1017/S0143385709000704} {\bibfield  {journal} {\bibinfo
  {journal} {Ergodic Theory and Dynamical Systems}\ }\textbf {\bibinfo {volume}
  {29}},\ \bibinfo {pages} {1881} (\bibinfo {year} {2009})}\BibitemShut
  {NoStop}\bibitem [{\citenamefont {Moussa}\ \emph {et~al.}(2023)\citenamefont {Moussa},
  \citenamefont {Xu}, \citenamefont {Yin}, \citenamefont {Galiffi},
  \citenamefont {Ra'di},\ and\ \citenamefont {Al{\`u}}}]{Moussa2023}  \BibitemOpen
  \bibfield  {author} {\bibinfo {author} {\bibfnamefont {H.}~\bibnamefont
  {Moussa}}, \bibinfo {author} {\bibfnamefont {G.}~\bibnamefont {Xu}}, \bibinfo
  {author} {\bibfnamefont {S.}~\bibnamefont {Yin}}, \bibinfo {author}
  {\bibfnamefont {E.}~\bibnamefont {Galiffi}}, \bibinfo {author} {\bibfnamefont
  {Y.}~\bibnamefont {Ra'di}},\ and\ \bibinfo {author} {\bibfnamefont
  {A.}~\bibnamefont {Al{\`u}}},\ }\bibfield  {title} {\bibinfo {title}
  {{Observation of temporal reflection and broadband frequency translation at
  photonic time interfaces}},\ }\href
  {https://doi.org/10.1038/s41567-023-01975-y} {\bibfield  {journal} {\bibinfo
  {journal} {Nat. Phys.}\ }\textbf {\bibinfo {volume} {19}},\ \bibinfo {pages}
  {863} (\bibinfo {year} {2023})}\BibitemShut {NoStop}\bibitem [{\citenamefont {Wang}\ \emph {et~al.}(2025)\citenamefont {Wang},
  \citenamefont {Shao}, \citenamefont {Chen}, \citenamefont {Chen},
  \citenamefont {Qian}, \citenamefont {Wu}, \citenamefont {Duan}, \citenamefont
  {Al{\'u}},\ and\ \citenamefont {Huang}}]{ElasticBeams2025}  \BibitemOpen
  \bibfield  {author} {\bibinfo {author} {\bibfnamefont {S.}~\bibnamefont
  {Wang}}, \bibinfo {author} {\bibfnamefont {N.}~\bibnamefont {Shao}}, \bibinfo
  {author} {\bibfnamefont {H.}~\bibnamefont {Chen}}, \bibinfo {author}
  {\bibfnamefont {J.}~\bibnamefont {Chen}}, \bibinfo {author} {\bibfnamefont
  {H.}~\bibnamefont {Qian}}, \bibinfo {author} {\bibfnamefont {Q.}~\bibnamefont
  {Wu}}, \bibinfo {author} {\bibfnamefont {H.}~\bibnamefont {Duan}}, \bibinfo
  {author} {\bibfnamefont {A.}~\bibnamefont {Al{\'u}}},\ and\ \bibinfo {author}
  {\bibfnamefont {G.}~\bibnamefont {Huang}},\ }\bibfield  {title} {\bibinfo
  {title} {{Experimental realization of temporal refraction and reflection in
  elastic beams}},\ }\href {https://doi.org/10.1038/s41467-025-64530-8}
  {\bibfield  {journal} {\bibinfo  {journal} {Nat. Commun.}\ }\textbf {\bibinfo
  {volume} {16}},\ \bibinfo {pages} {9520} (\bibinfo {year}
  {2025})}\BibitemShut {NoStop}\bibitem [{\citenamefont {Kim}\ \emph {et~al.}(2024)\citenamefont {Kim},
  \citenamefont {Chong},\ and\ \citenamefont {Daraio}}]{KimChongDaraio2024}  \BibitemOpen
  \bibfield  {author} {\bibinfo {author} {\bibfnamefont {B.~L.}\ \bibnamefont
  {Kim}}, \bibinfo {author} {\bibfnamefont {C.}~\bibnamefont {Chong}},\ and\
  \bibinfo {author} {\bibfnamefont {C.}~\bibnamefont {Daraio}},\ }\bibfield
  {title} {\bibinfo {title} {{Temporal Refraction in an Acoustic Phononic
  Lattice}},\ }\href {https://doi.org/10.1103/physrevlett.133.077201}
  {\bibfield  {journal} {\bibinfo  {journal} {Phys. Rev. Lett.}\ }\textbf
  {\bibinfo {volume} {133}},\ \bibinfo {pages} {077201} (\bibinfo {year}
  {2024})}\BibitemShut {NoStop}\bibitem [{\citenamefont {Hung}\ \emph {et~al.}(2013)\citenamefont {Hung},
  \citenamefont {Gurarie},\ and\ \citenamefont {Chin}}]{HungGurarieChin2013}  \BibitemOpen
  \bibfield  {author} {\bibinfo {author} {\bibfnamefont {C.-L.}\ \bibnamefont
  {Hung}}, \bibinfo {author} {\bibfnamefont {V.}~\bibnamefont {Gurarie}},\ and\
  \bibinfo {author} {\bibfnamefont {C.}~\bibnamefont {Chin}},\ }\bibfield
  {title} {\bibinfo {title} {{From Cosmology to Cold Atoms: Observation of
  Sakharov Oscillations in a Quenched Atomic Superfluid}},\ }\href
  {https://doi.org/10.1126/science.1237557} {\bibfield  {journal} {\bibinfo
  {journal} {Science}\ }\textbf {\bibinfo {volume} {341}},\ \bibinfo {pages}
  {1213} (\bibinfo {year} {2013})}\BibitemShut {NoStop}\bibitem [{\citenamefont {Sustaeta-Osuna}\ \emph {et~al.}(2025)\citenamefont
  {Sustaeta-Osuna}, \citenamefont {Garc{\'i}a-Vidal},\ and\ \citenamefont
  {Huidobro}}]{SustaetaOsuna2025}  \BibitemOpen
  \bibfield  {author} {\bibinfo {author} {\bibfnamefont {J.~E.}\ \bibnamefont
  {Sustaeta-Osuna}}, \bibinfo {author} {\bibfnamefont {F.~J.}\ \bibnamefont
  {Garc{\'i}a-Vidal}},\ and\ \bibinfo {author} {\bibfnamefont {P.~A.}\
  \bibnamefont {Huidobro}},\ }\bibfield  {title} {\bibinfo {title} {{Quantum
  Theory of Photon Pair Creation in Photonic Time Crystals}},\ }\href
  {https://doi.org/10.1021/acsphotonics.4c02293} {\bibfield  {journal}
  {\bibinfo  {journal} {ACS Photonics}\ }\textbf {\bibinfo {volume} {12}},\
  \bibinfo {pages} {1873} (\bibinfo {year} {2025})}\BibitemShut {NoStop}\bibitem [{\citenamefont {Eliasson}(1992)}]{Eliasson1992}  \BibitemOpen
  \bibfield  {author} {\bibinfo {author} {\bibfnamefont {L.~H.}\ \bibnamefont
  {Eliasson}},\ }\bibfield  {title} {\bibinfo {title} {{Floquet solutions for
  the 1-dimensional quasi-periodic Schr{\"o}dinger equation}},\ }\href
  {https://doi.org/10.1007/bf02097013} {\bibfield  {journal} {\bibinfo
  {journal} {Commun. Math. Phys.}\ }\textbf {\bibinfo {volume} {146}},\
  \bibinfo {pages} {447} (\bibinfo {year} {1992})}\BibitemShut {NoStop}\bibitem [{\citenamefont {Hou}\ and\ \citenamefont {You}(2012)}]{HouYou2012}  \BibitemOpen
  \bibfield  {author} {\bibinfo {author} {\bibfnamefont {X.}~\bibnamefont
  {Hou}}\ and\ \bibinfo {author} {\bibfnamefont {J.}~\bibnamefont {You}},\
  }\bibfield  {title} {\bibinfo {title} {{Almost reducibility and
  non-perturbative reducibility of quasi-periodic linear systems}},\ }\href
  {https://doi.org/10.1007/s00222-012-0379-2} {\bibfield  {journal} {\bibinfo
  {journal} {Invent. Math.}\ }\textbf {\bibinfo {volume} {190}},\ \bibinfo
  {pages} {209} (\bibinfo {year} {2012})}\BibitemShut {NoStop}\bibitem [{\citenamefont {Johnson}\ and\ \citenamefont
  {Moser}(1982)}]{JohnsonMoser1982}  \BibitemOpen
  \bibfield  {author} {\bibinfo {author} {\bibfnamefont {R.}~\bibnamefont
  {Johnson}}\ and\ \bibinfo {author} {\bibfnamefont {J.}~\bibnamefont
  {Moser}},\ }\bibfield  {title} {\bibinfo {title} {{The rotation number for
  almost periodic potentials}},\ }\href {https://doi.org/10.1007/bf01208484}
  {\bibfield  {journal} {\bibinfo  {journal} {Commun. Math. Phys.}\ }\textbf
  {\bibinfo {volume} {84}},\ \bibinfo {pages} {403} (\bibinfo {year}
  {1982})}\BibitemShut {NoStop}\end{thebibliography}
\end{document}